\documentclass[preprint,12pt]{elsarticle}

\usepackage{graphicx}
\usepackage{amssymb}
\usepackage{amsthm}
\usepackage{mathtools}
\usepackage{amsmath}
\usepackage{lineno}
\usepackage{mathrsfs}
\usepackage{booktabs, caption, makecell}
\usepackage{threeparttable}
\usepackage{algorithm}
\usepackage{algcompatible}
\usepackage{lipsum}
\usepackage{appendix}
\usepackage{float}

\usepackage[colorlinks,linkcolor=black]{hyperref}
\usepackage[nameinlink,capitalise]{cleveref}

\journal{Elsevier}

\begin{document}

\begin{frontmatter}
\title{Optimal sensor placement for reconstructing wind pressure field around buildings using compressed sensing}
\author[label1]{Xihaier Luo\corref{cor1}}
\ead{xluo@bnl.gov}

\author[label2]{Ahsan Kareem}

\author[label1]{Shinjae Yoo}

\cortext[cor1]{Corresponding author.}

\address[label1]{Computational Science Initiative, Brookhaven National Laboratory, Upton, NY 11973, United States.}
\address[label2]{NatHaz Modeling Laboratory, University of Notre Dame, Notre Dame, IN 46556, United States.}

\begin{abstract}
    Deciding how to optimally deploy sensors in a large, complex, and spatially extended structure is critical to ensure that the surface pressure field is accurately captured for subsequent analysis and design. In some cases, reconstruction of missing data is required in downstream tasks such as the development of digital twins. This paper presents a data-driven sparse sensor selection algorithm, aiming to provide the most information contents for reconstructing aerodynamic characteristics of wind pressures over tall building structures parsimoniously. The algorithm first fits a set of basis functions to the training data, then applies a computationally efficient QR algorithm that ranks existing pressure sensors in order of importance based on the state reconstruction to this tailored basis. The findings of this study show that the proposed algorithm successfully reconstructs the aerodynamic characteristics of tall buildings from sparse measurement locations, generating stable and optimal solutions across a range of conditions. As a result, this study serves as a promising first step toward leveraging the success of data-driven and machine learning algorithms to supplement traditional genetic algorithms currently used in wind engineering.
\end{abstract}


\end{frontmatter}

\section{Introduction}
\label{sec1}

As the world's population shifts to cities, the demand for high-rise buildings is rapidly increasing, as seen in the Pacific Rim region. Tall buildings exposed to wind experience wind-induced loads that create pressure on the building envelope, and their integral effects cause the structure to move in the dominant directions, namely along-wind, across-wind, and torsional \cite{kareem1992dynamic, gu2004across, kareem1985lateral}. The description of the pressure field around a building does not lend itself to a simple functional relationship with approach flow turbulence. As a result, calls for reliance on wind tunnel experiments have been made. These tests rely heavily on pressure taps connected to pressure sensors to monitor pressure fields over the building surface. \textcolor{black}{A basic question is where to deploy available sensors to accurately predict and reconstruct the structure of a wind pressure field from limited and noisy sensor outputs.}

\textcolor{black}{In fact, the optimal sensor placement problem has garnered considerable attention for a long time, as fast data acquisition, analysis, and decision in high-performance control for complex systems can be archived with a small number of measurements at limited locations.} In practice, the best locations for sensors in regular structures with simple geometry and a small number of degrees of freedom can be determined empirically using engineering judgment and past experience. However, for a complicated large-scale structure, a systematic and efficient approach is required because the solution space is far beyond the capabilities of hand calculation \cite{papadimitriou2004optimal, meo2005optimal, jiang2007pseudospectra, yi2011optimal, tan2020computational}. Mathematically, the goal is to find $m$ positions from a set of $n$ positions that maximize the information about the behaviors of a structural system:

\begin{equation}
\label{eq:combinational}
c=\frac{n !}{m !(n-m) !}
\end{equation}

where $n$ denotes the number of candidate positions, and $m$ denotes the number of sensors to be placed. It is worth noting that $m$ in equation \cref{eq:combinational} is not always a constant; rather, $m$ is often a parameter to be optimized (most of the time to be minimized) to minimize the total cost of sensors and their installation in a model's physically limited space while still providing adequate information on structural behaviors. \textcolor{black}{The current practice is based on wind tunnel facility experience, yet $c$ can be excessively huge for a typical modern tall building. Because a wind tunnel building model cannot be fully covered with sensors, a trade-off between needed information and expense must be made, necessitating automation in sensor position selection. This difficulty is exacerbated when attempting to locate a sensor in the presence of several wind-approaching angles.}


According to the literature, sensor locations are routinely chosen based on heuristics and intuition. For example, Yao et al. modified the parent selection and children reproduction schemes of genetic algorithms (GA) and applied them to optimal sensor placement on a large space structure for modal identification \cite{yao1993sensor}. Later, Liu et al. replaced the existing binary coding-based GA with the decimal two-dimension array coding method, resulting in a reduction in computational iterations \cite{liu2008optimal}. Another popular alternative is the Monkey algorithm (MA). For instance, Yi et al. developed an automatic technique for adjusting the MA's climb and watch–jump processes and applied it to the health monitoring of high-rise structures \cite{yi2015optimal}. Despite providing quick solutions to planning and scheduling problems, heuristic algorithms such as GA and MA do not always provide the best solution when compared to traditional optimization algorithms \cite{manohar2018data, erichson2020shallow, yang2021adaptive}. Their performance is highly dependent on the algorithmic parameters. Finding an effective set of parameters and iteration stopping criteria is case-dependent and difficult.

\textcolor{black}{This study focuses on a data-driven approach for reconstructing tall building aerodynamic characteristics. In comparison to existing sensor placement algorithms, the proposed approach makes use of cutting-edge decomposition-based sensing techniques, making it very efficient and effective in exploring a large space of candidate placements. A thorough investigation revealed that the proposed method not only scales well in terms of both the dimension of the wind pressure measurements and the number of sensors, but also delivers stable solutions in a wide range of wind conditions, such as different building features and wind attack angles.} 

\section{Methods}
\label{sec2}

\subsection{Compressed sensing}
\label{sec21}
Many natural signals are highly compressible, which means they can be well-approximated by a small number of non-zero coefficients in an appropriate basis \cite{brunton2022data}. A compressible signal $\mathbf{x} \in \mathbb{R}^{n}$ \textcolor{black}{(e.g., wind pressure distribution)} can be written mathematically as a sparse vector $\mathbf{s} \in \mathbb{R}^{n}$ on a new orthonormal basis of $\mathbf{\Psi} \in \mathbb{R}^{n \times n}$ such that:

\begin{equation}
\label{sec2_eq1}
\mathbf{x}=\mathbf{\Psi} \mathbf{s}
\end{equation}

If $\mathbf{s}$ in \cref{sec2_eq1} is a linear combination of only $k$ basis vectors, it is called $k$-sparse (exactly $k$ non-zero elements). Compressed sensing theory makes use of this idea in order to infer the sparse representations in a known transformed basis system $\mathbf{\Psi}$, where $k \ll n$. Without loss of generality, consider a set of observations $\mathbf{y} \in \mathbb{R}^{p}$ \textcolor{black}{(e.g., wind tunnel measurements)} and an observation matrix $\mathbf{C} \in \mathbb{R}^{p \times n}$ that satisfy

\begin{equation}
\label{sec2_eq2}
\mathbf{y}=\mathbf{C} \mathbf{x}=(\mathbf{C \Psi}) \mathbf{s}=\boldsymbol{\Theta} \mathbf{s}
\end{equation}

If $\mathbf{x}$ is sufficiently sparse in $\mathbf{\Psi}$ and the matrix $\boldsymbol{\Theta}$ follows the principle of restricted isometry, the search for $\mathbf{s}$ and the reconstruction of $\mathbf{x}$ can be expressed in an optimization format

\begin{equation}
\label{sec2_eq3}
\mathbf{s}=\arg \min _{\mathbf{s}^{\prime}}\left\|\mathbf{s}^{\prime}\right\|_{0} \quad \text { such that } \quad \mathbf{y}=\boldsymbol{\Theta} \mathbf{s}^{\prime}
\end{equation}

\cref{sec2_eq3} entails a difficult combinatorial search. In practice, the goal of compressed sensing is to find the $l_1$-norm of a sparsest vector that is consistent with $\mathbf{y}$

\begin{equation}
\label{sec2_eq4}
\mathbf{s}=\arg \min _{\mathbf{s}^{\prime}}\left\|\mathbf{s}^{\prime}\right\|_{1} \quad \text { such that } \quad \mathbf{y}=\boldsymbol{\Theta} \mathbf{s}^{\prime}
\end{equation}

where $\|\mathbf{s}\|_{1}=\sum_{k=1}^{n}\left|s_{k}\right|$. By doing so, the combinatorially difficult problem in the nonconvex $l_0$ minimization is bypassed by relaxing to a convex $l_1$ minimization \cite{candes2006stable, donoho2006compressed, candes2008introduction}.

\subsection{Data-driven sparse sensor placement}
\label{sec22}
While compressed sensing employs random measurements to reconstruct high-dimensional unknown data from a universal basis $\mathbf{\Psi} \in \mathbb{R}^{n \times n}$, data-driven sparse sensor placement, as discussed in this paper, collects available information about a signal from observed samples to build a tailored basis $\mathbf{\Psi}_r \in \mathbb{R}^{n \times r}$ for the respective signal and thus identify optimal sensor placements for low-loss reconstruction \cite{manohar2018data, brunton2022data}. Herein, \textcolor{black}{the wind pressure signal} in \cref{sec2_eq1} can be rewritten as an unknown linear combination of basis coefficients:

\begin{equation}
\label{sec2_eq5}
\mathbf{x}=\sum_{k=1}^{r} \psi_{k} a_{k}=\mathbf{\Psi}_{r} \mathbf{a}
\end{equation}

where vector $\mathbf{a} \in \mathbb{R}^{r}$ represents mode amplitudes of $\mathbf{x}$ in basis $\mathbf{\Psi}$. Similarly, the reduction process described in \cref{sec2_eq2} can be expressed as:

\begin{equation}
\label{sec2_eq6}
\mathbf{y}=\mathbf{C x}=\left(\mathbf{C} \boldsymbol{\Psi}_{r}\right) \mathbf{a}=\boldsymbol{\Theta a}
\end{equation}

The main challenge is to design an incoherent measurement matrix $\mathbf{C}$ that allows for the identification of the optimal $p$ observations $\mathbf{y}$ for accurately reproducing the signal $\mathbf{x}$. In other words, rows of $\mathbf{C}$ not correlated with columns $\psi$ of $\boldsymbol{\Psi}_{r}$. The data-driven sparse sensor placement algorithm addresses this issue by solving

\begin{equation}
\label{sec2_eq7}
\mathbf{C}^{\star}=\underset{\mathbf{C} \in \mathbb{R}^{p \times n}}{\arg \min }\left|\mathbf{x}-\mathbf{\Psi}(\mathbf{C \Psi})^{\dagger} \mathbf{y}\right|_{2}^{2}
\end{equation}

where $\dagger$ denotes the Moore-Penrose pseudoinverse. Mathematically, \cref{sec2_eq7} is the Moore-Penrose pseudoinverse of \cref{sec2_eq6}. It is assumed that $\mathbf{C}^{\star}$ is a mostly sparse subset selection operator composed of rows of the identity, with nonzero entries indicating the chosen measurements. Specifically, $\mathbf{C}$ is constrained to have the following structure given a $p$ sensor budget and $n$ candidates state components:

\begin{equation}
\label{sec2_eq8}
\mathbf{C}=\left[\begin{array}{llll}
\mathbf{e}_{\gamma_{1}}^{\top} & \mathbf{e}_{\gamma_{2}}^{\top} & \cdots & \mathbf{e}_{\gamma_{p}}^{\top}
\end{array}\right]
\end{equation}

where $\mathbf{e}_{\gamma_{j}}^{\top}$ is the canonical basis vector with a unit entry in the $j^{th}$ component and zeros everywhere else. \textcolor{black}{In practice, $\mathbf{e}_{\gamma_{j}}^{\top}$ denotes the location of a given sensor.} As a result, each row of $\mathbf{C}$ observes from a single spatial location, which corresponds to the sensor location:

\begin{equation}
\label{sec2_eq9}
\mathbf{y}=\mathbf{C x}=\left[x_{\gamma_{1}}, x_{\gamma_{2}}, \ldots, x_{\gamma_{p}}\right]^{\top}
\end{equation}

For a well-defined linear inverse problem, the index with cardinality $|\gamma|=p$ and additionally number of sensors $n \geq r$ of $\mathbf{\Psi}$ is designated by $\gamma \in \mathbb{N}^{p}$ \cite{candes2006robust, gilbert2010sparse}.

\subsubsection{Implementation note 1: Basis functions}
\label{sec221}
The measurement matrix is an important aspect of compressed sensing. In the context of data-driven sparse sensor placement, this means that the basis functions $\mathbf{\Psi}$ in \cref{sec2_eq7} will be replaced by $\mathbf{\Psi}_r$, which is built from the training data $\mathbf{X}^{tr}$. There are various approaches to building $\mathbf{\Psi}_r$ depending on the available information.

\textit{Identity basis:} Compressed sensing is ideal for recovering a high dimensional signal with unknown content by employing random measurements on a global scale. In other words, the raw measurement data is used directly and without modification: $\boldsymbol{\Psi}_{r}=\mathbf{X}^{tr}$. Because there is no low-rank approximation of the data, there is no information loss. This, however, comes at the expense of a longer computation time.

\textit{Random Projection basis:} Even in cases with insufficient explicit information about the signal, the computation of the identity basis can be speeded up by projecting the input data onto a randomly generated matrix, that is, multiplying measurements with random Gaussian vectors to project them to a new space $\boldsymbol{\Psi}_{r}=\mathbf{G} \mathbf{X}^{tr}$, where the entries $\mathbf{G} \in \mathbb{R}^{2 p \times m}$ are drawn from a Gaussian density function with mean zero and variance $1/n$. This basis is also known as the Random Projection basis \cite{li2006very, dasgupta2013experiments}.

\textcolor{black}{\textit{SVD/POD basis:}} If information about the type of signal is available (for example, whether it is a turbulent velocity field or an image of a cat), it is possible to design optimized sensors that are tailored to the specific signals of interest. \textcolor{black}{Dimensionality reduction techniques can be used to extract dominant features from a training data set of representative examples.} These low-rank features, derived from data patterns, aid in the development of identifying optimal sensors locations. We used the singular value decomposition (SVD) in this study because of its numerical robustness and efficiency in extracting dominant patterns from low-dimension data \cite{fowler2009compressive, halko2011finding}. \textcolor{black}{SVD can also be used to solve the proper orthogonal decomposition (POD), which is widely used in wind data analysis and reconstruction.} For a matrix $\mathbf{X}^{tr} \in \mathbb{C}^{n \times n}$, the SVD basis is computed as:

\begin{equation}
\label{sec2_eq10}
\mathbf{X}^{tr}=\mathbf{U} \boldsymbol{\Sigma} \mathbf{V}^{T}=\boldsymbol{\Psi} \boldsymbol{\Sigma} \mathbf{V}^{T} \approx \boldsymbol{\Psi}_{r} \boldsymbol{\Sigma}_{r} \mathbf{V}_{r}^{T}
\end{equation}

where the matrices $\boldsymbol{\Psi}_{r}$ and $\mathbf{V}_{r}$ contain the first $r$ columns of left and right singular vectors, respectively. In this study, we will examine and investigate the three types of basis functions mentioned above in the context of reconstructing aerodynamic wind pressure fields.

\subsubsection{Implementation note 2: QR pivoting}
\label{sec222}
The first implementation focuses on determining the best-fitting basis for the training data. The second implementation note we provided aims to use a sparse sensor selection algorithm that is both computationally efficient and flexible. To achieve the goal, the greedy matrix QR factorization with column pivoting is used to effectively determine the resulting optimal sensor locations that minimize the reconstruction error \cite{brunton2022data, higham2000qr}. Specifically, QR factorization breaks a matrix $\mathbf{A} \in \mathbb{R}^{m \times n}$ down into a unitary matrix $\mathbf{Q}$, an upper-triangular matrix $\mathbf{R}$, and a column permutation matrix $\mathbf{C}$ defined in \cref{sec2_eq8} such that

\begin{equation}
\label{sec2_eq11}
\mathbf{A} \mathbf{C}^{T}=\mathbf{Q R}
\end{equation}

QR column pivoting increases the volume of the submatrix built from the pivoted columns by choosing a new pivot column with the highest two-norm and then subtracting from each other column its orthogonal projection onto the pivot column. Pivoting increases the volume of the submatrix by enforcing a diagonal dominance structure.

\begin{equation}
\label{sec2_eq12}
\sigma_{i}^{2}=\left|r_{i i}\right|^{2} \geq \sum_{j=i}^{k}\left|r_{j k}\right|^{2} ; \quad 1 \leq i \leq k \leq m
\end{equation}

This works because the product of diagonal entries is also the product of matrix volume.

\begin{equation}
\label{sec2_eq13}
|\operatorname{det} \mathbf{A}|=\prod_{i} \sigma_{i}=\prod_{i}\left|r_{i i}\right|
\end{equation}

Also, the oversampled case $p > r$ can be solved using $\boldsymbol{\Psi}_{r} \boldsymbol{\Psi}_{r}^{T}$'s pivoted QR factorization, with the column pivots chosen from $n$ candidate state-space locations based on the observation that

\begin{equation}
\label{sec2_eq14}
\operatorname{det} \boldsymbol{\Theta}^{T} \boldsymbol{\Theta}=\prod_{i=1}^{r} \sigma_{i}\left(\boldsymbol{\Theta} \boldsymbol{\Theta}^{T}\right)
\end{equation}

As a result, QR factorization combined with column pivoting produces $r$ column indices (corresponding to sensor locations) that best sample the $r$ basis modes (columns)

\begin{equation}
\label{sec2_eq15}
\mathbf{\Psi}_{\mathbf{r}}^{T} \mathbf{C}^{T}=\mathbf{Q R}
\end{equation}

Because the pivots columns represent the sensors, the QR-factorization produces a hierarchical list of all $n$ pivots, with the first $p$ pivots optimized for $\mathbf{\Psi}_{r}$ reconstruction. This means that all wind pressure taps are ranked based on their information content in the aerodynamic reconstruction based on the data. When spatial input data, such as interpolation or model results, is used, all gridded input data cells are ranked based on their information content. As a result, it is possible to make recommendations for the placement of additional sensors in areas with a high information content \cite{manohar2018data, halko2011finding}.

\subsection{Method overview}
\label{sec23}

\begin{figure}[b!]
    \centering
    \includegraphics[width=1.0\textwidth]{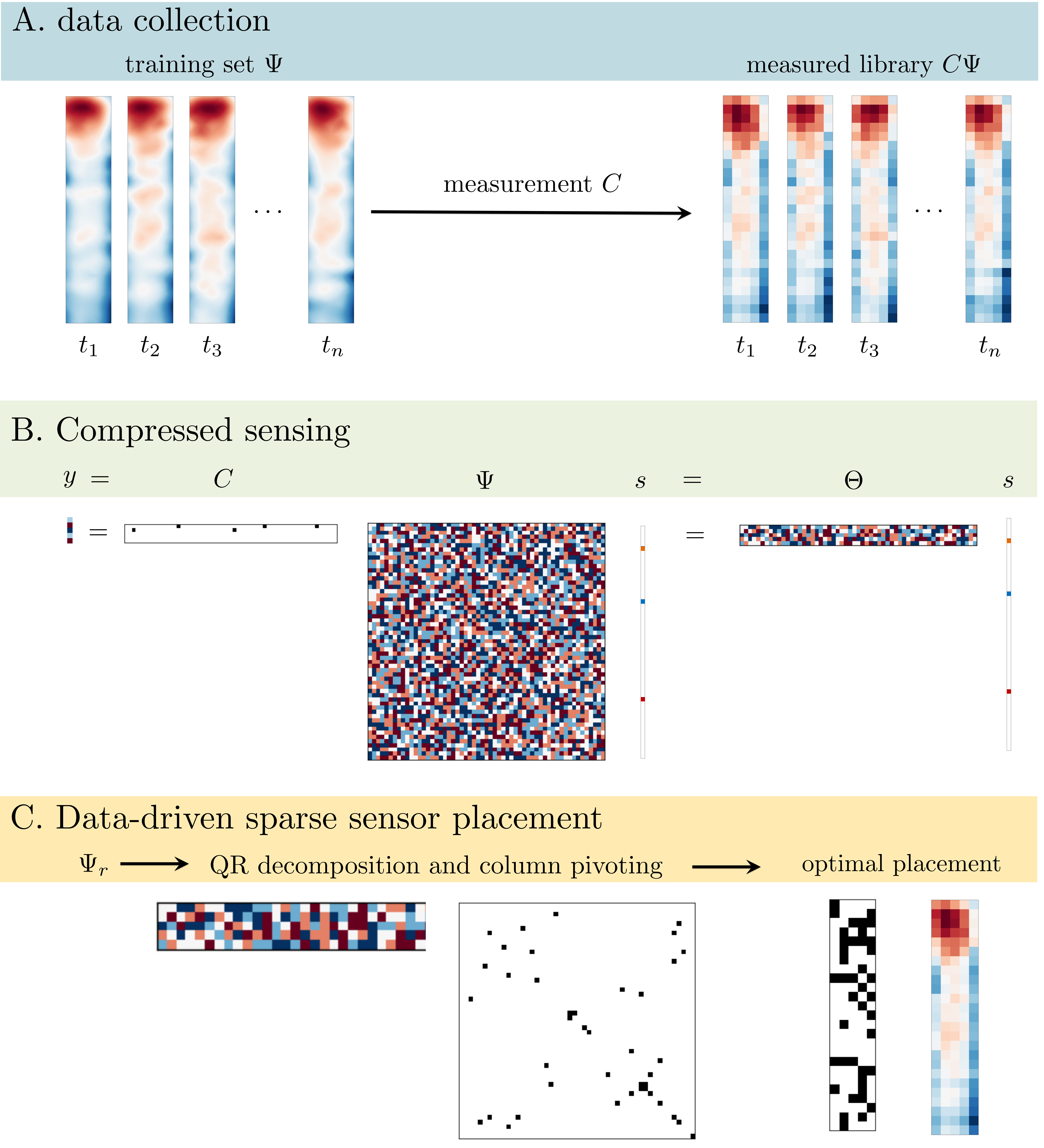}
    \caption{\textcolor{black}{Flow diagram of compressed sensing-based optimal sensor placement for reconstructing wind pressure fields.}}
    \label{fig:f0}
\end{figure}

\textcolor{black}{The goal of the proposed method is to estimate and reconstruct a detailed wind pressure field from limited and possibly noise-contaminated measurements. To achieve the goal, we first need to create the training data set and build the measurement operator $C$ using the wind tunnel data. Following the preceding compressive sensing techniques, the solution provides the sparsest solution to an underdetermined linear problem. \cref{fig:f0} depicts compressed sensing graphically (\cref{sec2_eq1} and \cref{sec2_eq2}). Next, a tailored basis function for our training data is constructed using a singular value decomposition (\cref{sec2_eq10}). We next execute a QR decomposition with column pivoting to produce pivots corresponding to our chosen sensor positions (\cref{sec2_eq15}). Finally, a relaxed convex optimization problem is performed to estimate sparse coefficients (\cref{sec2_eq7}), ensuring that the reconstructed data is consistent with the noisy measurements.}

\section{Experimental database}
\label{sec3}

\begin{figure}[H]
    \centering
    \includegraphics[width=0.9\textwidth]{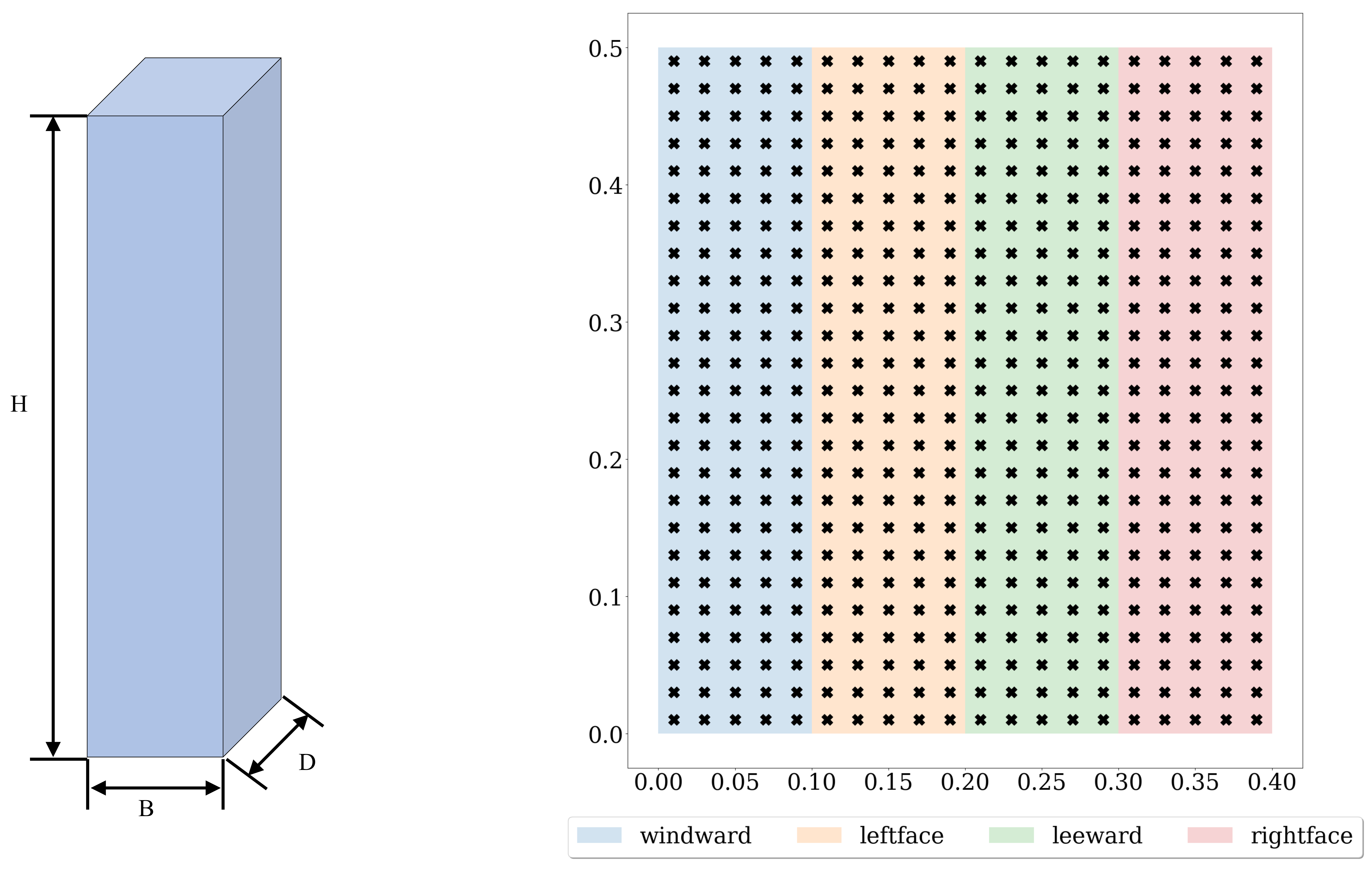}
    \caption{Schematic of the scaled building model and the distribution of pressure taps.}
    \label{fig:f1}
\end{figure}

The Tokyo Polytechnic University (TPU) Wind Engineering Information Center provides a comprehensive wind pressure database derived from a series of wind tunnel tests on a wide range of buildings \cite{TPU}. Specifically, the tests were carried out in a boundary layer wind tunnel with a test section 1.2 m wide by 1.0 m high, and the scaled model (1/400) had an identical cross-section with a width of 0.1 m. Turbulence-generating spires, roughness elements, and a carpet on the upstream floor of the wind tunnel's test section were used to simulate the atmospheric boundary layer. The database was built using various wind profiles. Most experiments were carried out under a power-law wind profile with an exponent of 1/4. The mean wind speed at the top of the building was 11 m/s and the turbulence intensity profiles is in accordance with the category III (suburban terrain). As shown in \cref{fig:f1}, a total of 500 pressure taps were used to collect data at a sampling frequency of 1000 Hz for a sample period of 32.768 secs. Pressure taps were evenly distributed on building surfaces with 0.02 m row spacing and 5 columns on each face, with model dimensions of 100 mm $\times$ 100 mm $\times$ 500 mm (breadth $\times$ depth $\times$ height).

\section{Results and Discussion}
\label{sec4}

\subsection{Different choices of basis}
\label{sec41}

\begin{figure}[H]
    \centering
    \includegraphics[width=0.9\textwidth]{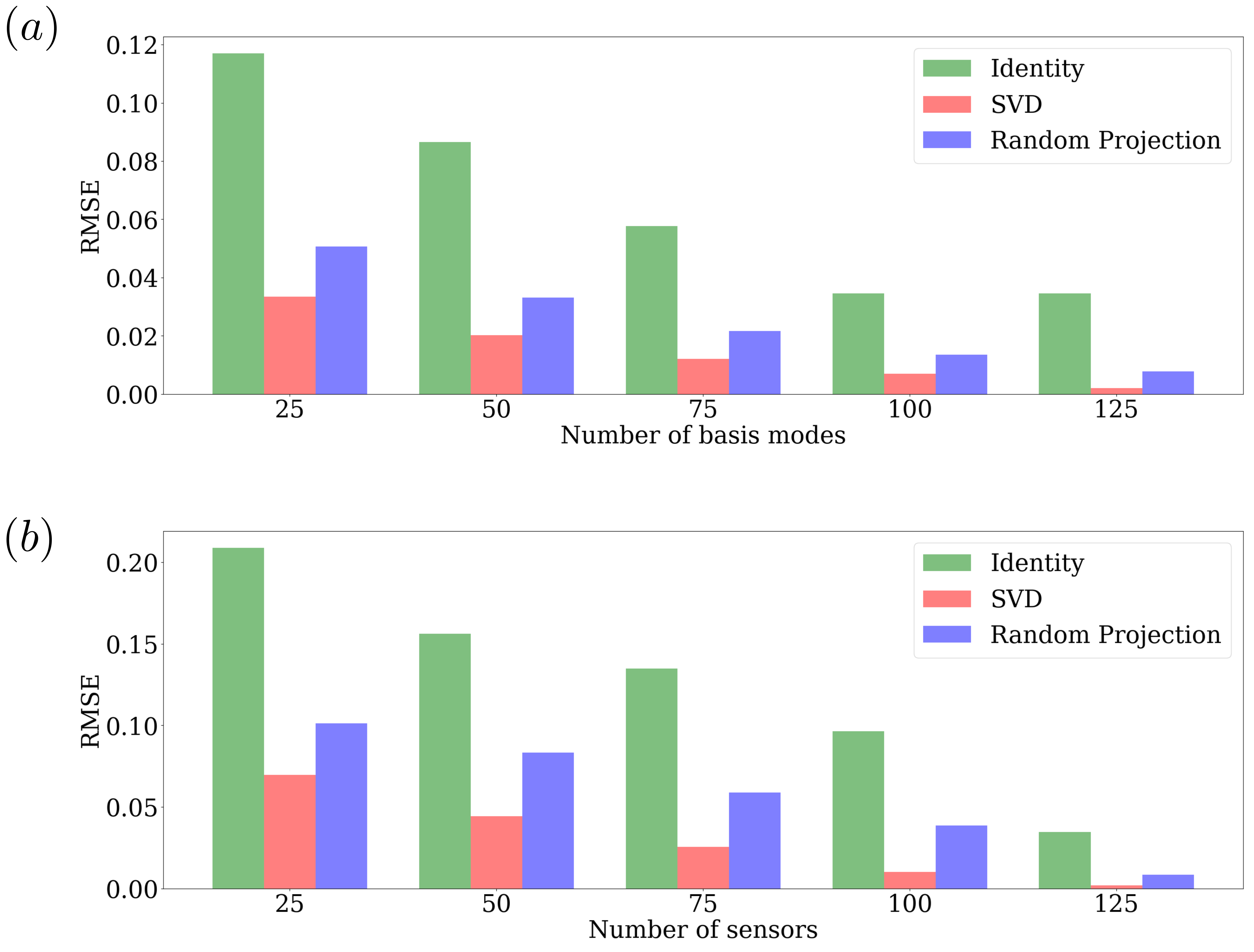}
    \caption{Singular value spectrum.}
    \label{fig:f2}
\end{figure}

It is critical to choose a good basis for accurately reconstructing the aerodynamics of wind-excited tall buildings. In this section, we start with specifying a set of parameter values to test for the aforementioned three types of tailored basis (identity basis, random projection, and SVD basis). Because the number of sensors and basis modes interacts, it is necessary to independently determine the effects of different types of the tailored basis on these two factors \cite{manohar2018data, erichson2020shallow}. As a result, we designed two experiments. In the first set, we kept the number of sensors constant at 125 and investigate how the number of basis modes used affects the reconstruction error. In the second set, the reconstruction error for a fixed number of basis modes (125) as the number of sensors varies is investigated. The computing results (\cref{fig:f2}) show that the Identity basis performs the worst, while the SVD basis consistently outperforms the other two types of basis. Meanwhile, in these experiments, we have five reduction stages: 20$\%$, 40$\%$, 60$\%$, 80$\%$, and 100$\%$. The reduction ratio is calculated using the total number of installed sensors, i.e., 125. In general, as the number of sensors and basis modes increases, the reconstruction error decreases. Though the presented results are based on windward data, similar results have been observed in the other three building faces. For consistency reasons and based on the grid search results, the SVD basis will be used as the default basis in the following experiments.

\subsection{Varying the number of sensors/basis modes}
\label{sec42}
\begin{figure}[H]
    \centering
    \includegraphics[width=0.7\textwidth]{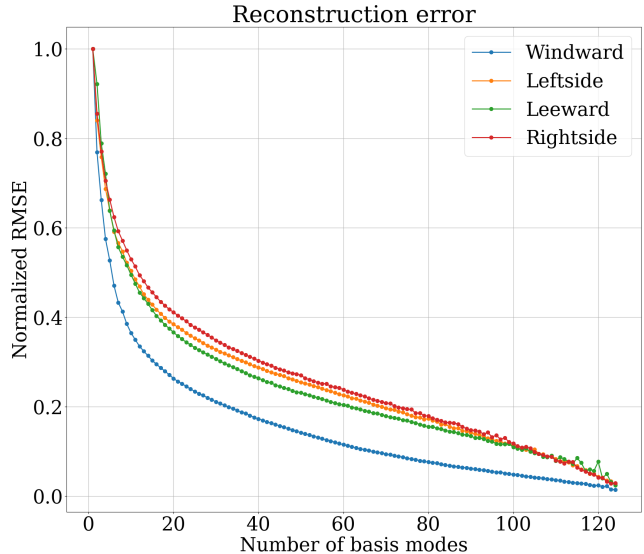}
    \caption{Reconstruction error versus the number of basis modes.}
    \label{fig:f3}
\end{figure}

\begin{figure}[H]
    \centering
    \includegraphics[width=0.7\textwidth]{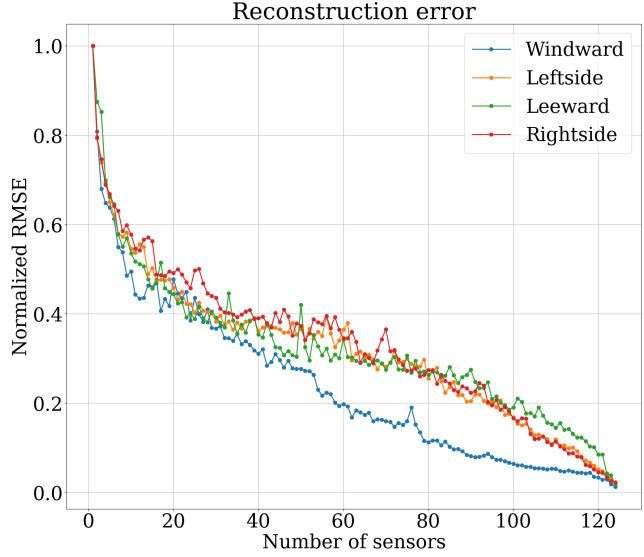}
    \caption{Reconstruction error versus the number of sensors.}
    \label{fig:f4}
\end{figure}

Choosing the fewest sensors and basis modes for reconstruction is an essential requirement of an optimal sensor placement. Because the proposed data-driven sensor placement optimization relies on low-rank structure in the data and involves an inherent trade-off between the number of sensors/basis modes and reconstruction accuracy. \cref{fig:f3} and \cref{fig:f4} show the results of experiments with varying numbers of basis modes and sensors. In all experiments, 7000 snapshots are used to train the SVD modes and different sensor/basis mode sets, and these different sensor sets are used to reconstruct a set of 3000 validation snapshots that were not used for training features. The computed results show that the reconstruction error decreases rapidly at first, indicating that the proposed method can provide accurate reconstruction with a limited number of sensors/basis modes. Four building surfaces exhibit a similar trend in terms of the number of basis modes, with less than 20 basis modes providing a faithful reconstruction. The windward data, in particular, promotes efficient characterization. When 17 basis modes are used, the proposed algorithm performs $30\%$ better in the windward scenario than in the other three surfaces. Such findings are consistent with the inherent aerodynamics of wind passing around a bluff prism, where more complex wind-structure interactions (e.g., separation and reattachment phenomenon) are frequently detected on two sides and leeward facets.

Another point to note is that the relationship between reconstruction performance and the number of sensors is less defined than the number of basis modes. The noise could be causing the fluctuations. Measurements of real-world data are frequently contaminated by sensor noise. The SVD-based sensor selection is theoretically optimal for estimation with measurements corrupted by zero-mean Gaussian white noise. Developing an accurate noise model for these non-stationary and non-Gaussian signals remains a challenge. As a result, the elbow of the reconstruction error curve down and to the left exhibits some fluctuations, particularly when the number of sensors is small (See \cref{fig:f4}). As more sensors are added, the model can better capture intrinsic noise and stabilize performance. When more than $60\%$ of the sensors are used, in our case 80 sensors, the fluctuation becomes very small.

\subsection{Reconstruction performance: temporal perspective}
\label{sec43}

\begin{figure}[H]
    \centering
    \includegraphics[width=0.85\textwidth]{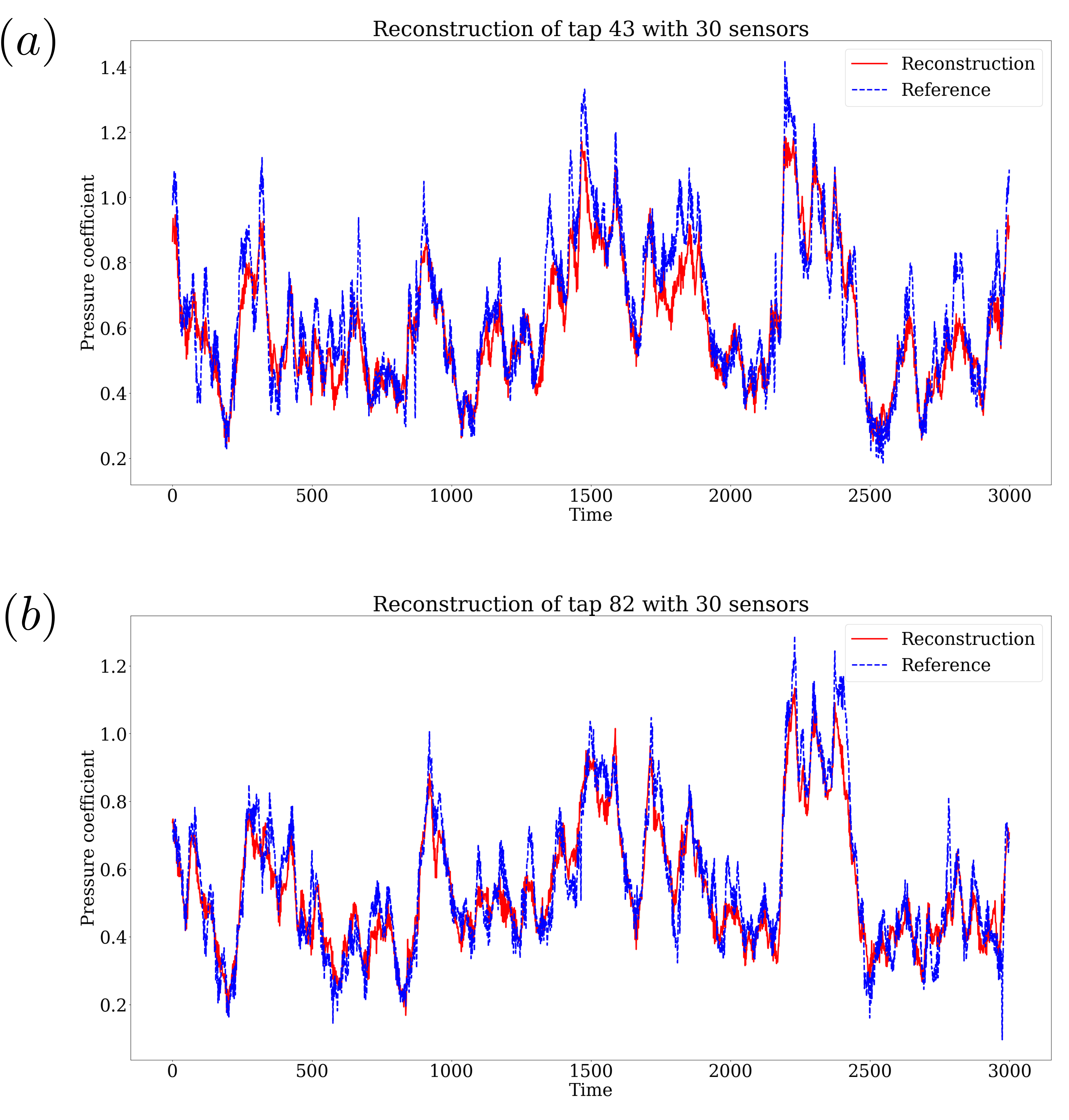}
    \caption{Reconstructed time history of randomly selected pressure taps field using $25\%$ sensors.}
    \label{fig:f5_p1}
\end{figure}

\begin{figure}[H]
    \centering
    \includegraphics[width=0.85\textwidth]{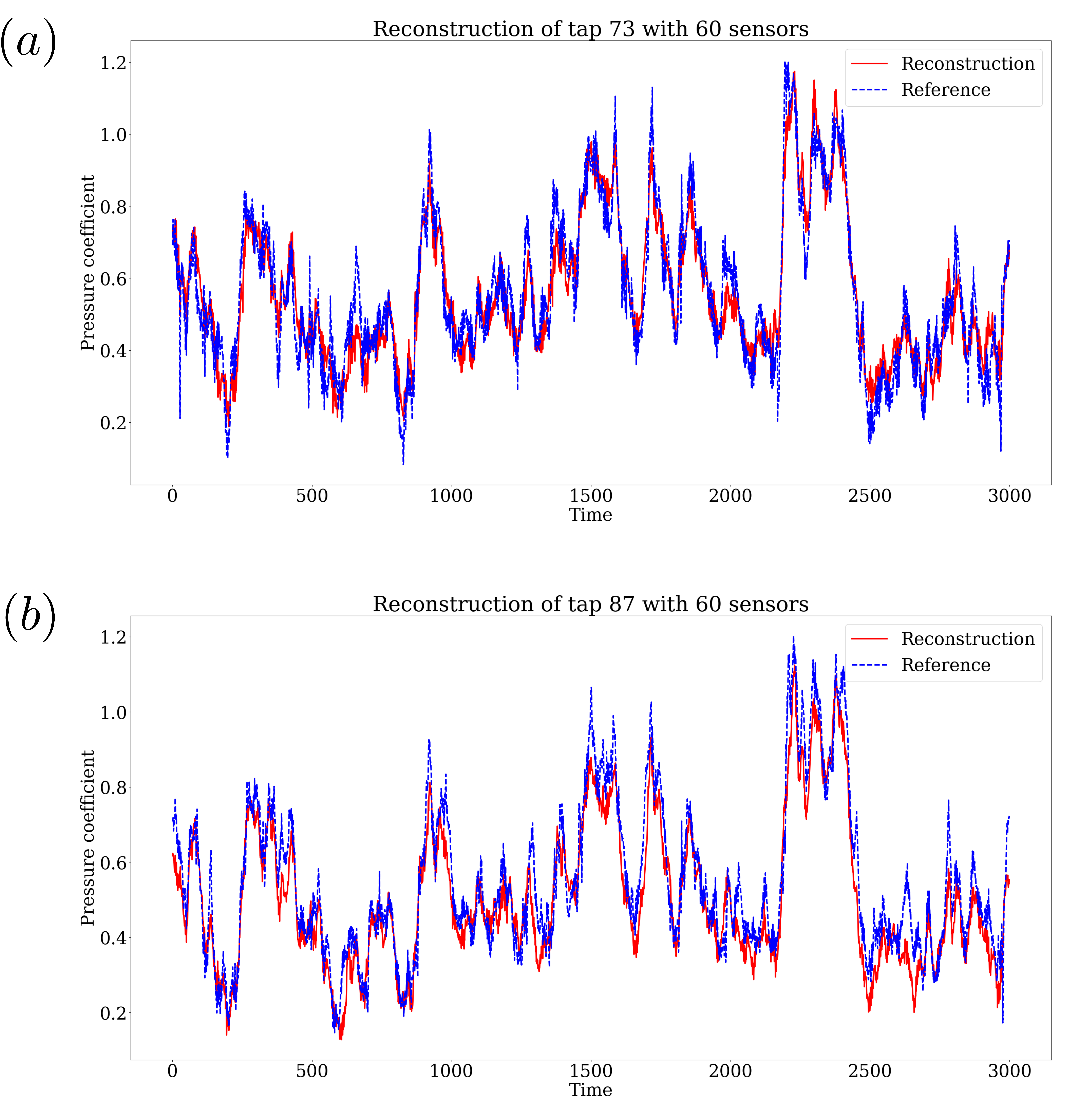}
    \caption{Reconstructed time history of randomly selected pressure taps field using $50\%$ sensors.}
    \label{fig:f5_p2}
\end{figure}

\begin{figure}[H]
    \centering
    \includegraphics[width=0.85\textwidth]{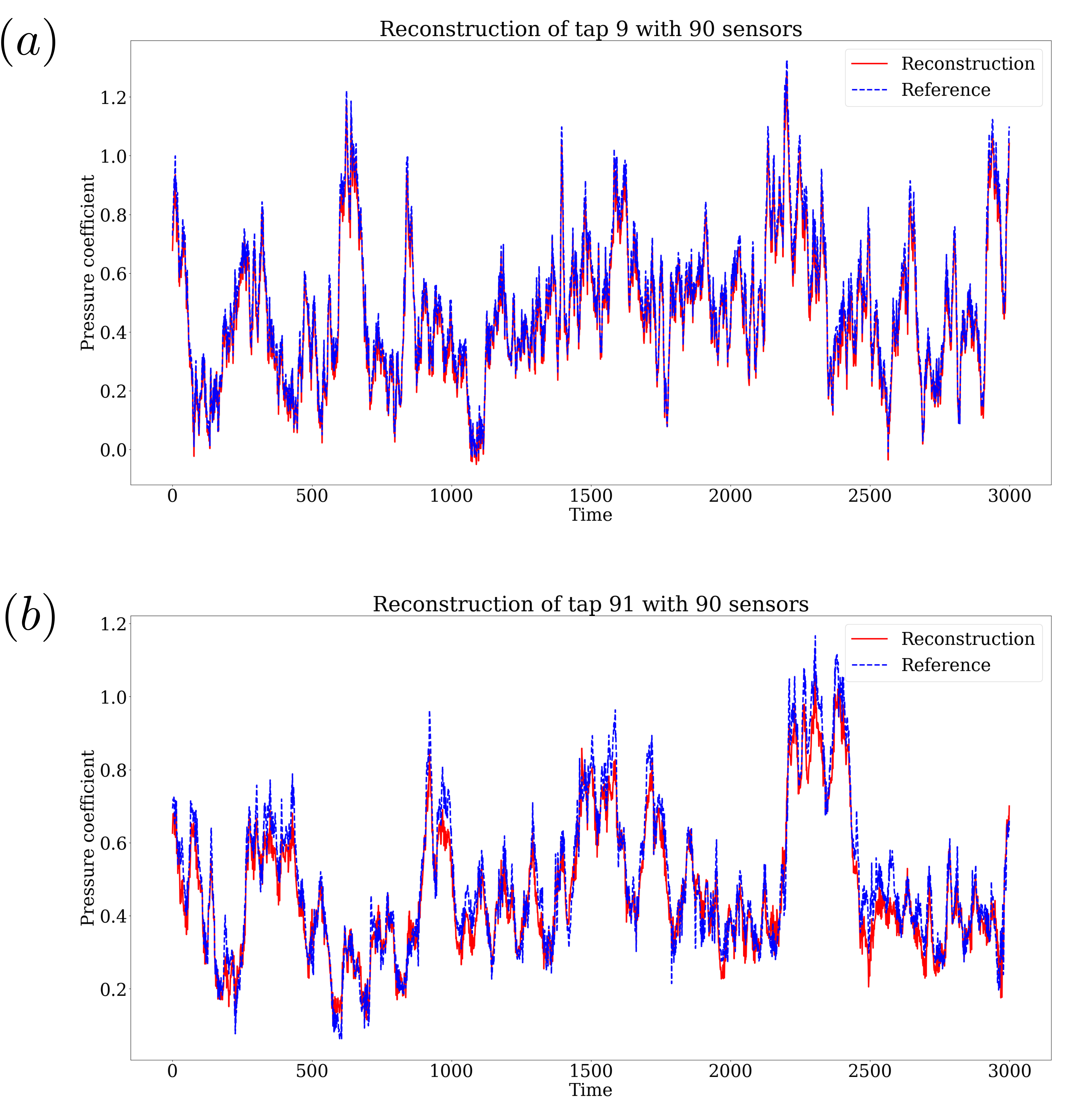}
    \caption{Reconstructed time history of randomly selected pressure taps using $70\%$ sensors.}
    \label{fig:f5_p3}
\end{figure}

The proposed algorithm's reconstruction performance using a different number of sensors is evaluated using wind pressure monitoring data from the testing dataset. Note the experimental data from wind tunnel tests is pre-processed as the non-dimensional wind pressure coefficient. Usually, wind pressure is defined by the pressure coefficient $C_{p}=\frac{P_{x}-P_{0}}{\rho U_{h}^{2} / 2}$, where $P_{x}$ is the static pressure at a given point on the building facade, $P_{0}$ is the static reference pressure at freestream, $\rho U_{h}^{2} / 2$ is the dynamic pressure at freestream, $\rho$ is the air density, and $U_{h}$ is the wind speed, which is frequently measured at the building height h in the undisturbed flow upstream. \cref{fig:f5_p1}-\cref{fig:f5_p3} show the reconstructed wind pressure data in comparison to the measured wind pressure data. Because each surface has 125 pressure taps, we chose two at random for demonstration. Clearly, the reconstructed data is highly consistent with the measured data, despite the fact that only a few sensors were used (30, 50, and 70 in the presented experiments). The reconstruction performance of four building surfaces is comparable overall, particularly when 50 percent of sensors, i.e., 60 sensors, are available. The algorithm captures the extremes better in the case of 60 sensors. And as the number of sensors increases (up to 125), the performance in matching extreme excursions in data improves. In summary, the proposed algorithm can effectively reconstruct wind pressure data using a small number of sensors, opening up a new avenue for addressing difficult problems such as optimal sensor placement, filling in missed data from faulty sensors, and so on.

\subsection{Optimal sensor locations}
\label{sec44}
\begin{figure}[H]
    \centering
    \includegraphics[width=0.9\textwidth]{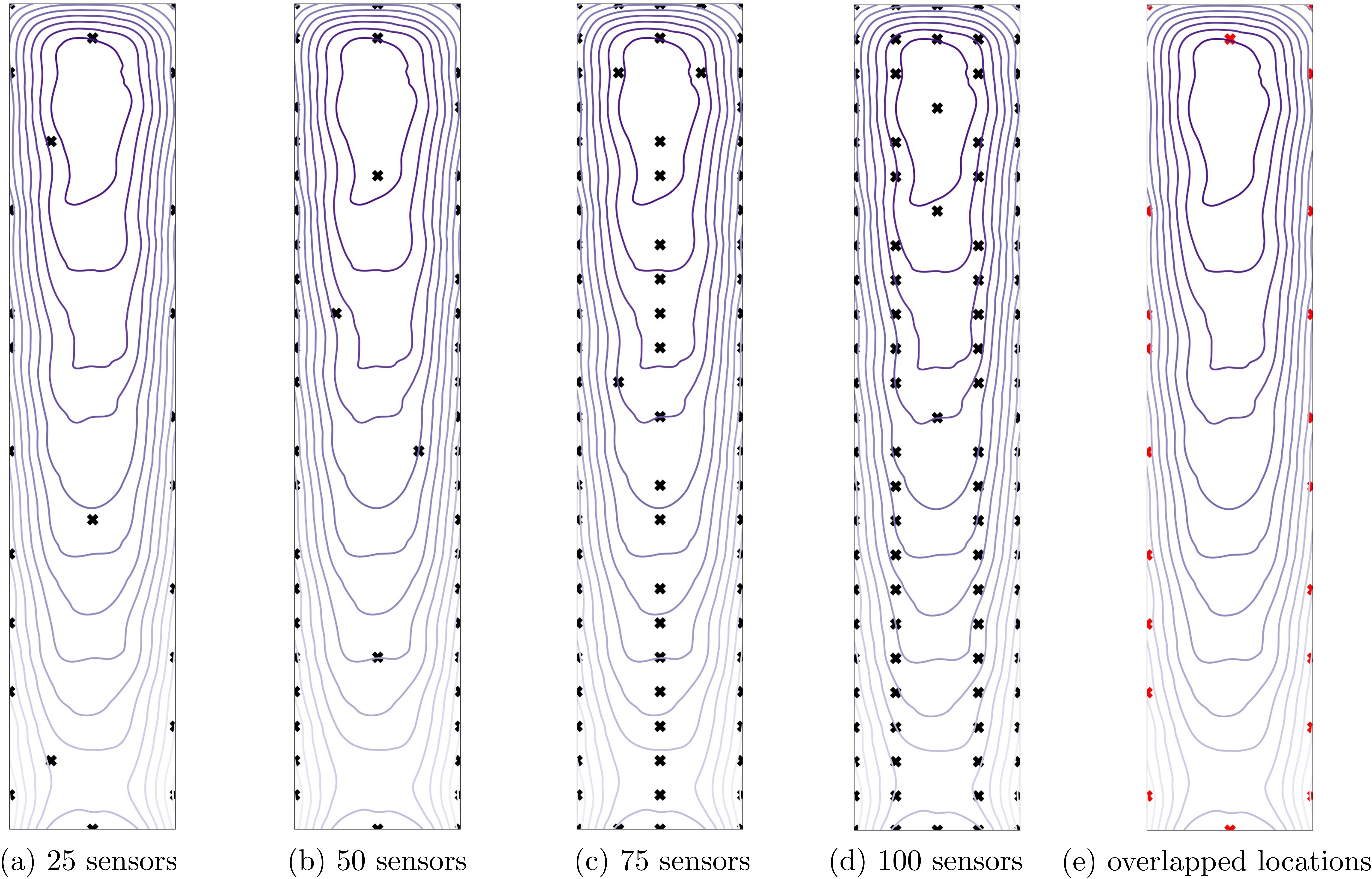}
    \caption{Optimal sensor locations: windward scenario.}
    \label{fig:f6}
\end{figure}

\begin{figure}[H]
    \centering
    \includegraphics[width=0.9\textwidth]{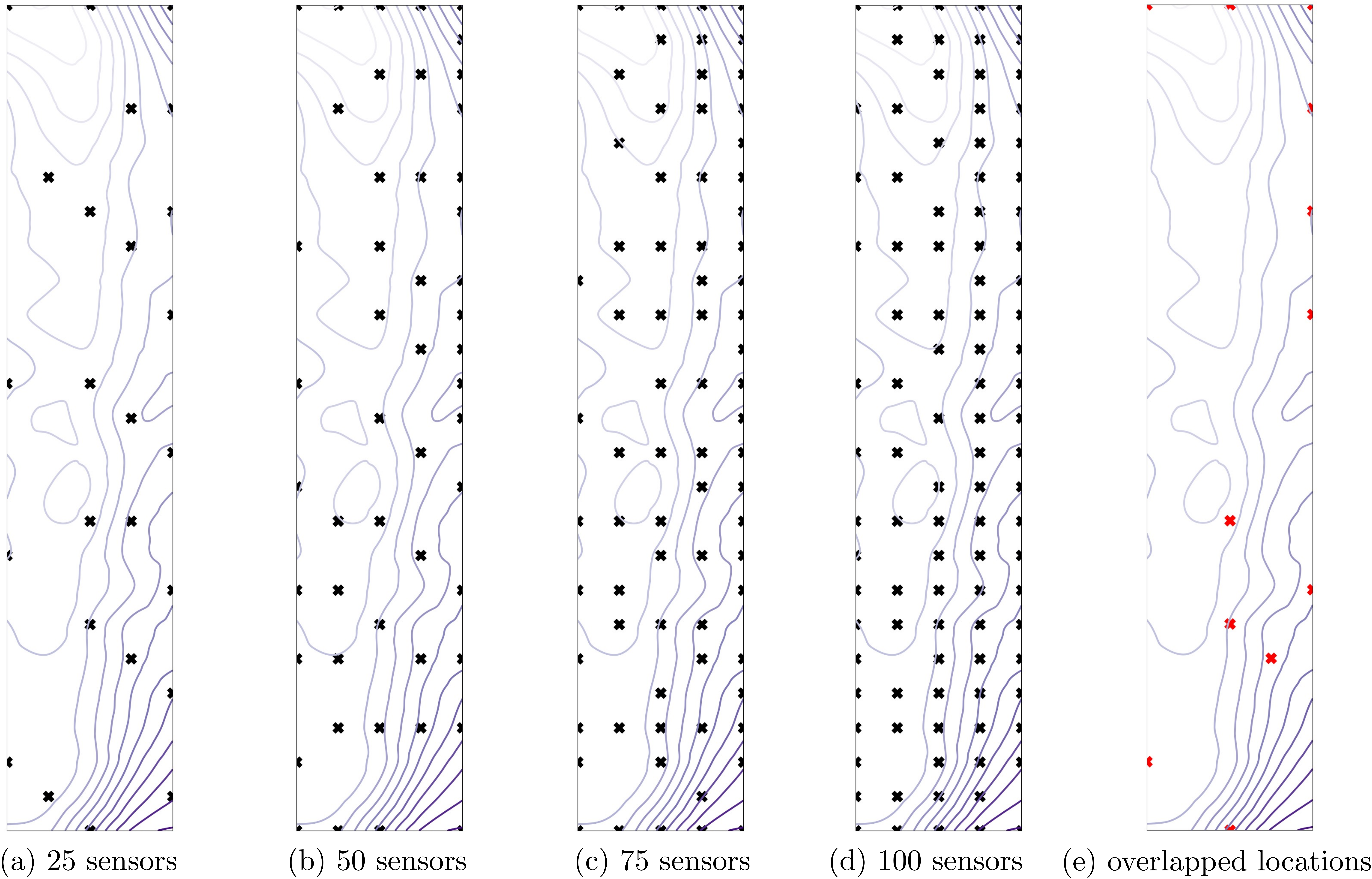}
    \caption{Optimal sensor locations: leftside scenario.}
    \label{fig:f7}
\end{figure}

\begin{figure}[H]
    \centering
    \includegraphics[width=0.9\textwidth]{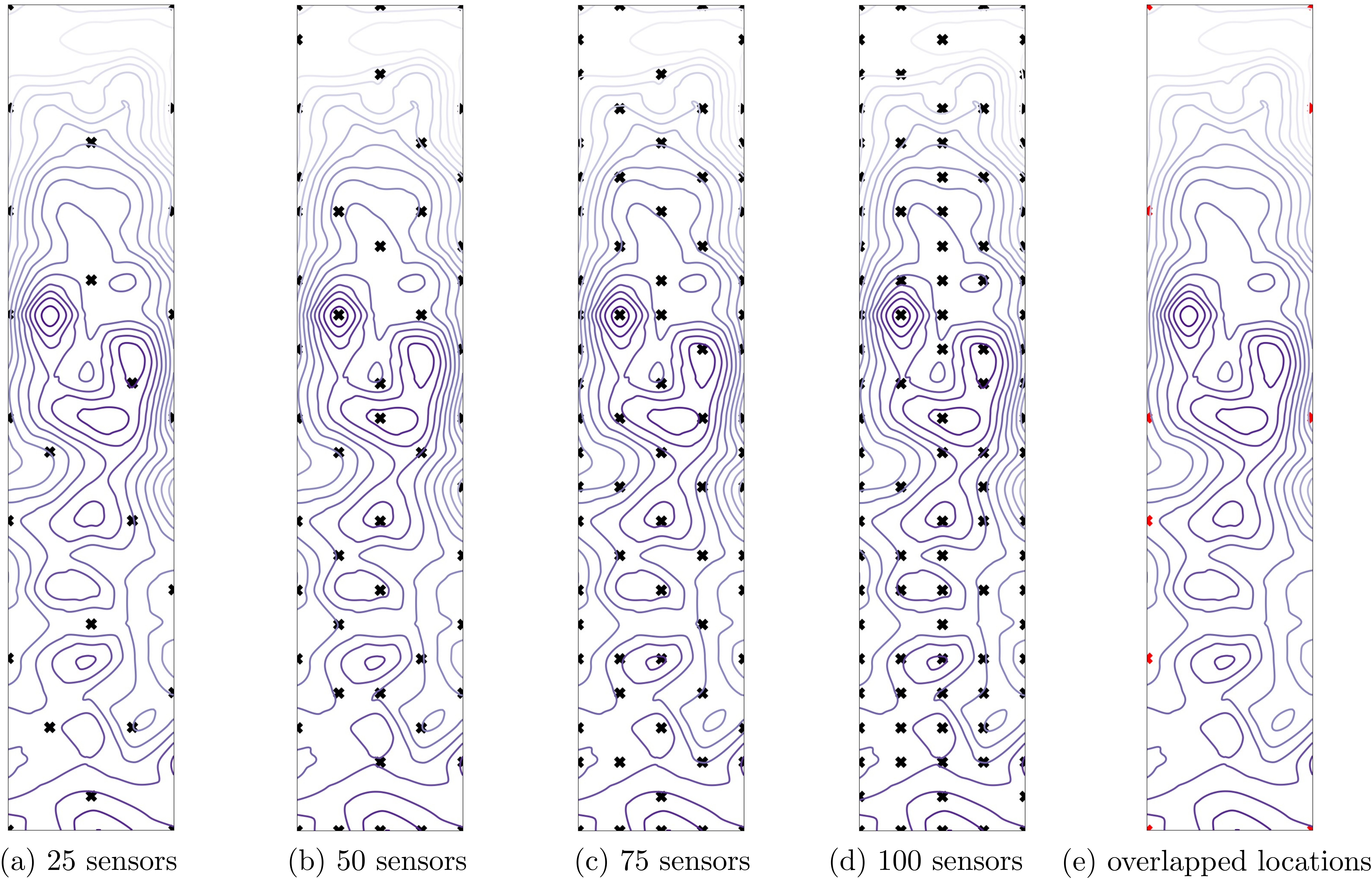}
    \caption{Optimal sensor locations: leeward scenario.}
    \label{fig:f8}
\end{figure}

\begin{figure}[H]
    \centering
    \includegraphics[width=0.9\textwidth]{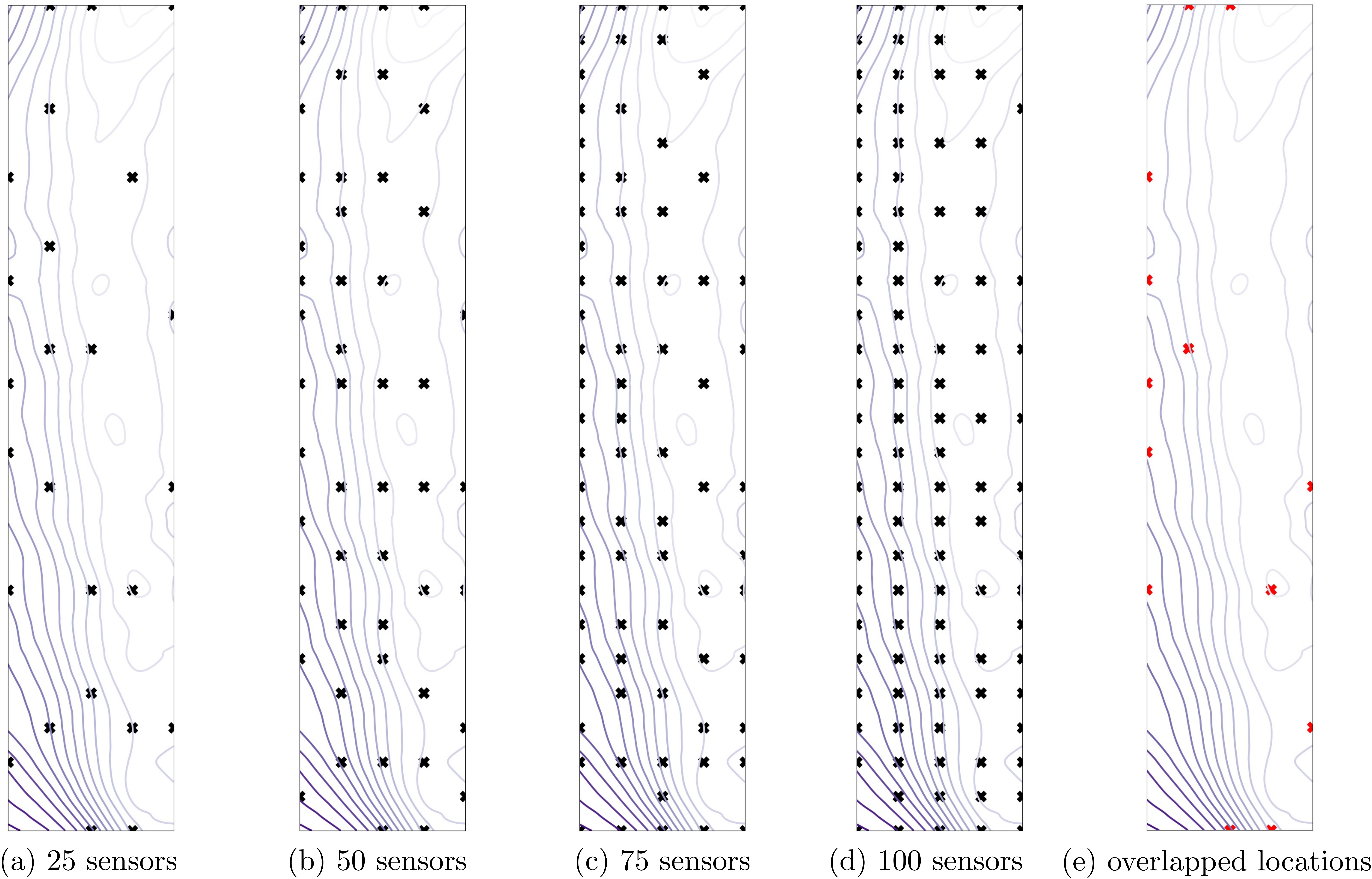}
    \caption{Optimal sensor locations: rightside scenario.}
    \label{fig:f9}
\end{figure}

In this section, we examine the reconstruction performance from sensor spatial distribution, that is, the optimal sensor locations. We run experiments with an increasing number of sensors, 25 (20$\%$), 50 (40$\%$), 75 (60$\%$), and 100 (80$\%$), for various building surfaces. In the last column of \cref{fig:f6}-\cref{fig:f9}, we extracted the overlapped positions, which are locations that have been identified in all experiment trials. The mean pressure distribution contour of a square building model is shown as the background in these figures. A few observations about the results are made here.

First, the optimal windward sensor locations are relatively symmetrical, which agrees with the mean pressure distribution of buildings with square sections. Large-scale coherent structures, such as spanwise vortex, tip vortex, and horseshoe vortex, naturally exist in the flow field around high-rise buildings. Experimental evidence shows the interaction of these vortex structures controls the fluctuating feature of pressure \cite{kareem1992dynamic, tamura1997proper, zhao2017effects}. As a result, wind pressure sensors should be placed near two side edges (See \cref{fig:f6}). Furthermore, the windward contour shows that the pressure coefficient reaches its maximum value at approximately 4/5 of the height. And, when more than $50\%$ of sensors are available, the proposed data-driven algorithm automatically distributes some sensors around the maximum value region. In parallel, the lowest value of wind pressure is observed near the ground at the windward edge. Again, the optimal sensors identified include one in the center of the ground line and a few near the bottom corners (See the middle two sensors in the \cref{fig:f6} (e).). These two findings suggest that the optimal sensor locations identified are physically interpretable in terms of first reconstructing the large-scale pressure patterns and then restoring the minimum and maximum value regions.

Second, wake fluctuation and vortex shedding are known to influence the aerodynamics of the two side facets. The separated shear layers produced by the leading edges of the windward surface usually roll up to form vortices. The vortex shedding then effects pressure on the building model's side surfaces. Interestingly, the identified sensor locations are not symmetric \cite{kareem1984pressure, kim2021pressure}. As a result of vortex shedding not appearing symmetrically due to the strong suction effect, an asymmetrical distribution on the two side surfaces emerges. The background mean pressure contours confirm this. The right side had more turbulent interactions, which resulted in more local patterns. For example, more small contours are detected in the middle of the prism. As a result, a greater number of sensors have been distributed to the corresponding region to ensure reconstruction quality (See the last column of \cref{fig:f7} and \cref{fig:f9}).

\begin{figure}[H]
    \centering
    \includegraphics[width=0.85\textwidth]{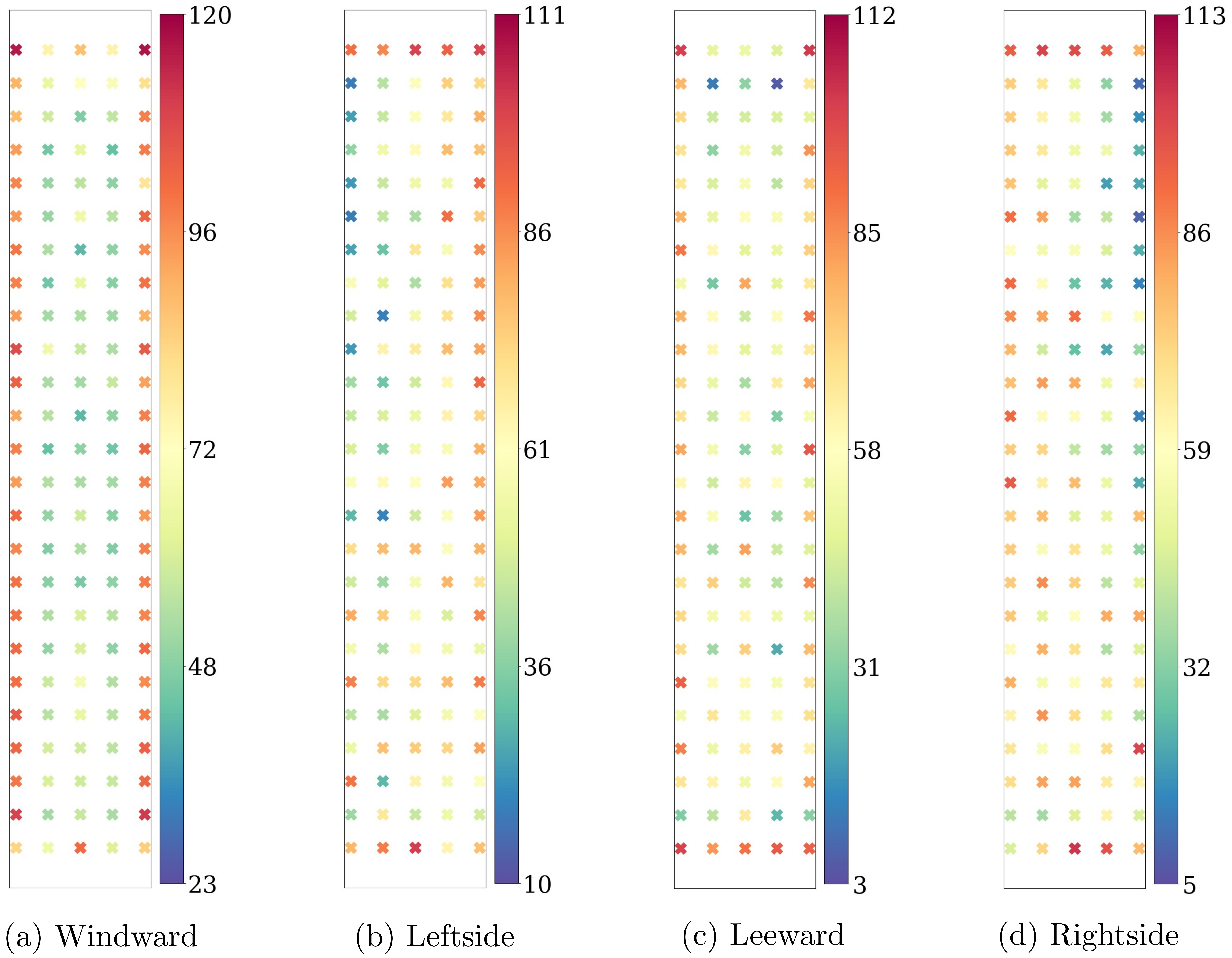}
    \caption{\textcolor{black}{The wind pressure taps' importance ranking, where importance is defined as the accumulated count of the sensor location being identified. Higher numbers indicate a higher ranking in terms of information content and importance in reconstructing wind pressure data.}}
    \label{fig:f10}
\end{figure}

The ranking of the wind pressure taps is shown in \cref{fig:f10}. The numbers represent the accumulated count of positions identified during an iteration from one sensor to the full rank. \textcolor{black}{Specifically, we run a series of optimal sensor placement experiments. The input is all 125 sensors for each building surface, and the output is the number of sensors we want to keep, which can range from 1 to 125, resulting in a total of 125 experiments. In theory, this means that the lower bound for the importance value is 1 (the experiment in which all of the sensors are kept) and the upper bound is 125 (a sensor identified in each of the 125 experiments). As a result, the higher the value, the more critical the sensor location is for analyzing and reconstructing measured wind pressure data.} Overall, the ranking demonstrates that the most important sensors are located in positions that contain information about larger pressure patterns, such as mean, variance, skewness, and kurtosis. Striped and micro-scale patterns identified by the dynamic mode decomposition method, for example, are labeled with a lower rank \cite{luo2021dynamic, zhou2021higher}. This finding is consistent with physical nature because these patterns are local and have higher frequency contents, contributing less to data reconstruction at the macro-scale.

\subsection{Varying the wind angles}
\label{sec45}

\begin{figure}[H]
    \centering
    \includegraphics[width=0.95\textwidth]{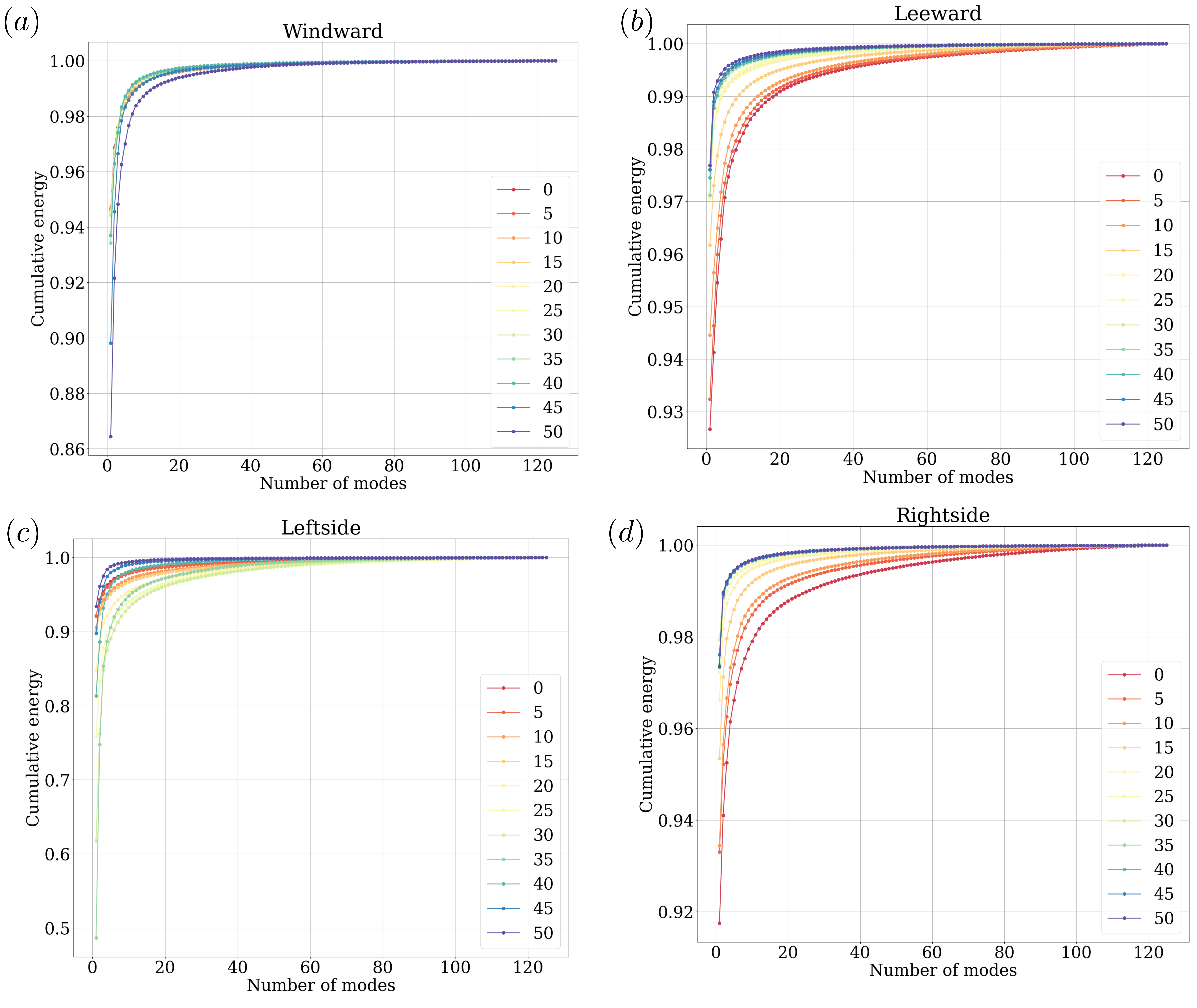}
    \caption{\textcolor{black}{Cumulative energy of SVD/POD modes. The analysis is performed with wind angles ranging from $0^{\circ}$ to $50^{\circ}$ degrees every $5^{\circ}$}}
    \label{fig:f11}
\end{figure}

When wind flows around buildings, characteristics of wind separation, vortex shedding, and wake effect are distinct under different wind directions, and wind pressure coefficients follow accordingly. As a result, when reconstructing wind pressure data, it is critical to consider the effects of wind direction. We investigate the effect of wind angles ranging from $0^{\circ}$ to $50^{\circ}$ degrees every $5^{\circ}$. However, before conducting sensor analysis, we must ensure that our previous assumptions about the number of basis and modes, as well as the type of basis, are applicable to new wind angles. \textcolor{black}{We used proper orthogonal decomposition (POD) to better align the introduced algorithm with current baseline methods for wind data analysis and reconstruction. The SVD and POD are related in that the SVD is an extension of eigendecomposition to rectangular matrices, and POD can be viewed as a decomposition formalism for which the SVD is one approach to determining its solution. Specifically, $\lambda_{POD} = \lambda_{SVD}^2$ connects the singular values and the POD eigenvalues \cite{taira2017modal}. \cref{fig:f11} shows the ranked POD modes based on their energy content. The POD eigenvalue distribution depicts the wind energy captured by the accumulated POD modes as a percentage of the reconstructed total turbulent kinetic energy. The energy of POD modes decreases dramatically with mode number, as shown by the line plots. The first POD mode represents the mean distribution of wind pressures. With a few more modes, the accumulated decomposed mode can contain relatively large fluctuating energy. The overall trend is very similar to the findings in \cref{sec42},} implying that the SVD basis applies to new wind angles. Furthermore, we observe that as the wind angle increases, the SVD value decreases more rapidly. This finding is consistent with existing experimental findings and means that a smaller number of basis is supported for sensor placement, making the proposed algorithm even more robust and efficient \cite{meng2018sensitivity}.

\begin{figure}[H]
    \centering
    \includegraphics[width=0.9\textwidth]{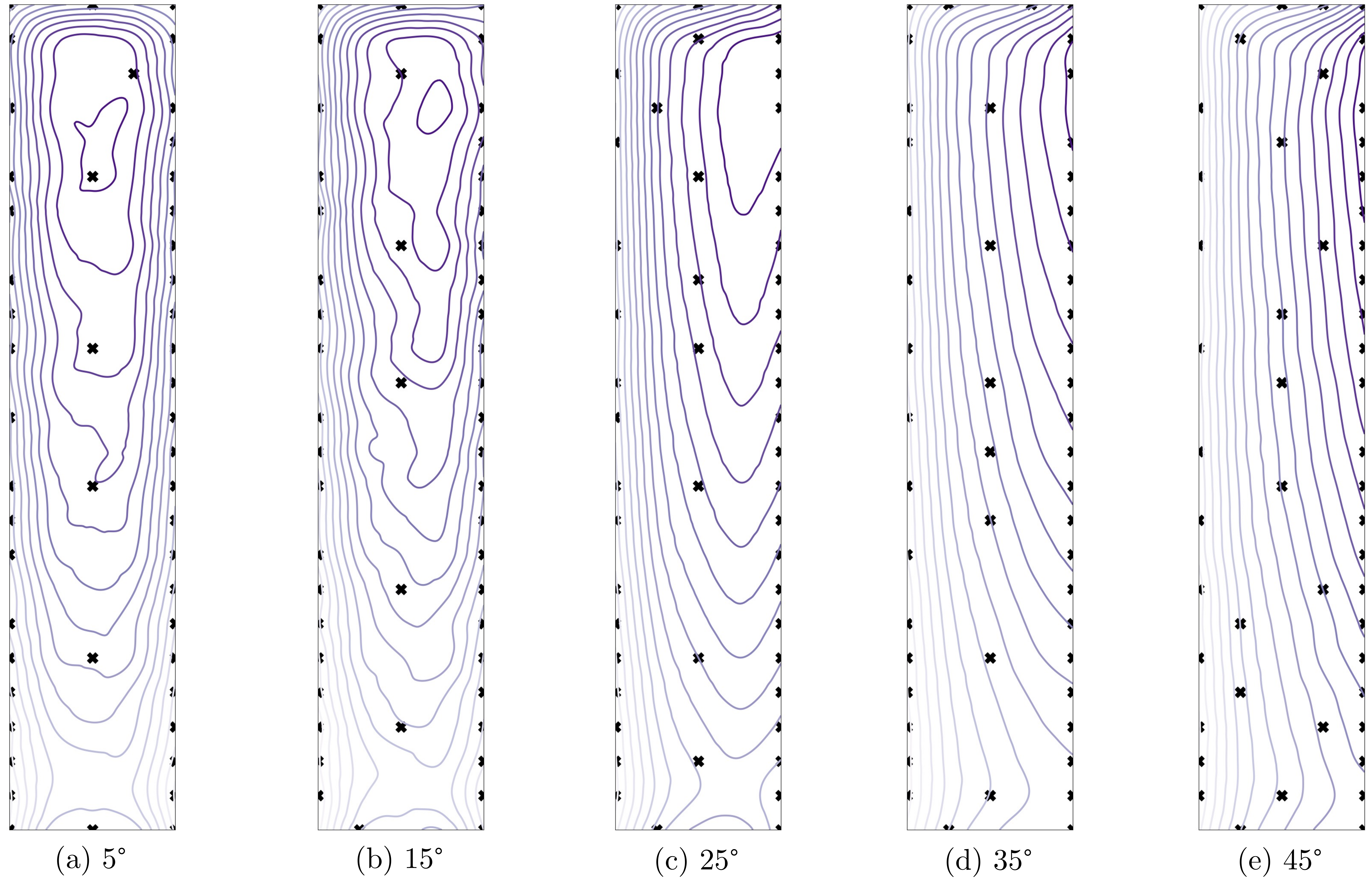}
    \caption{Optimal sensor locations: windward scenario.}
    \label{fig:f12}
\end{figure}

\begin{figure}[H]
    \centering
    \includegraphics[width=0.9\textwidth]{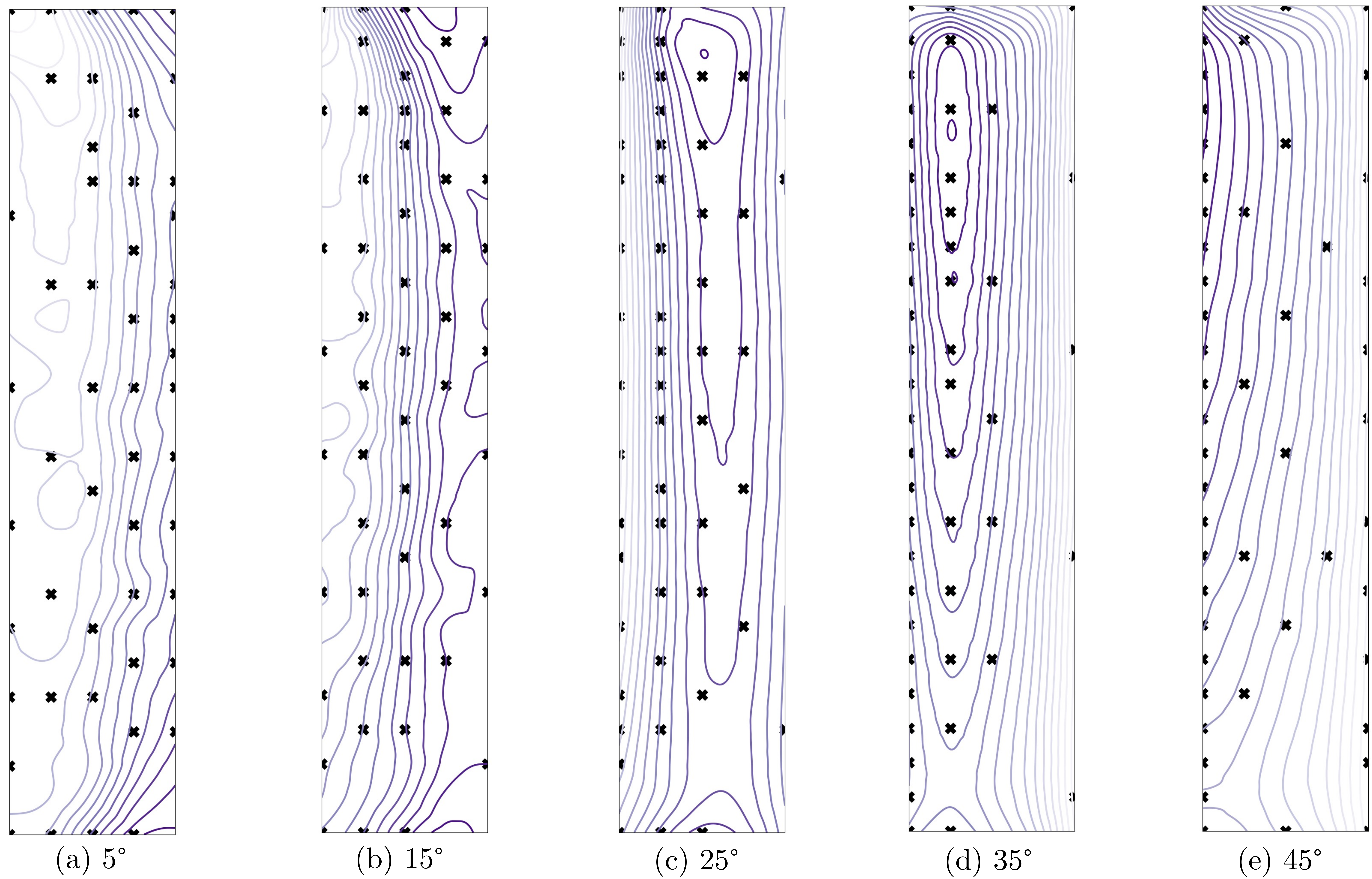}
    \caption{Optimal sensor locations: leftside scenario.}
    \label{fig:f13}
\end{figure}

\begin{figure}[H]
    \centering
    \includegraphics[width=0.9\textwidth]{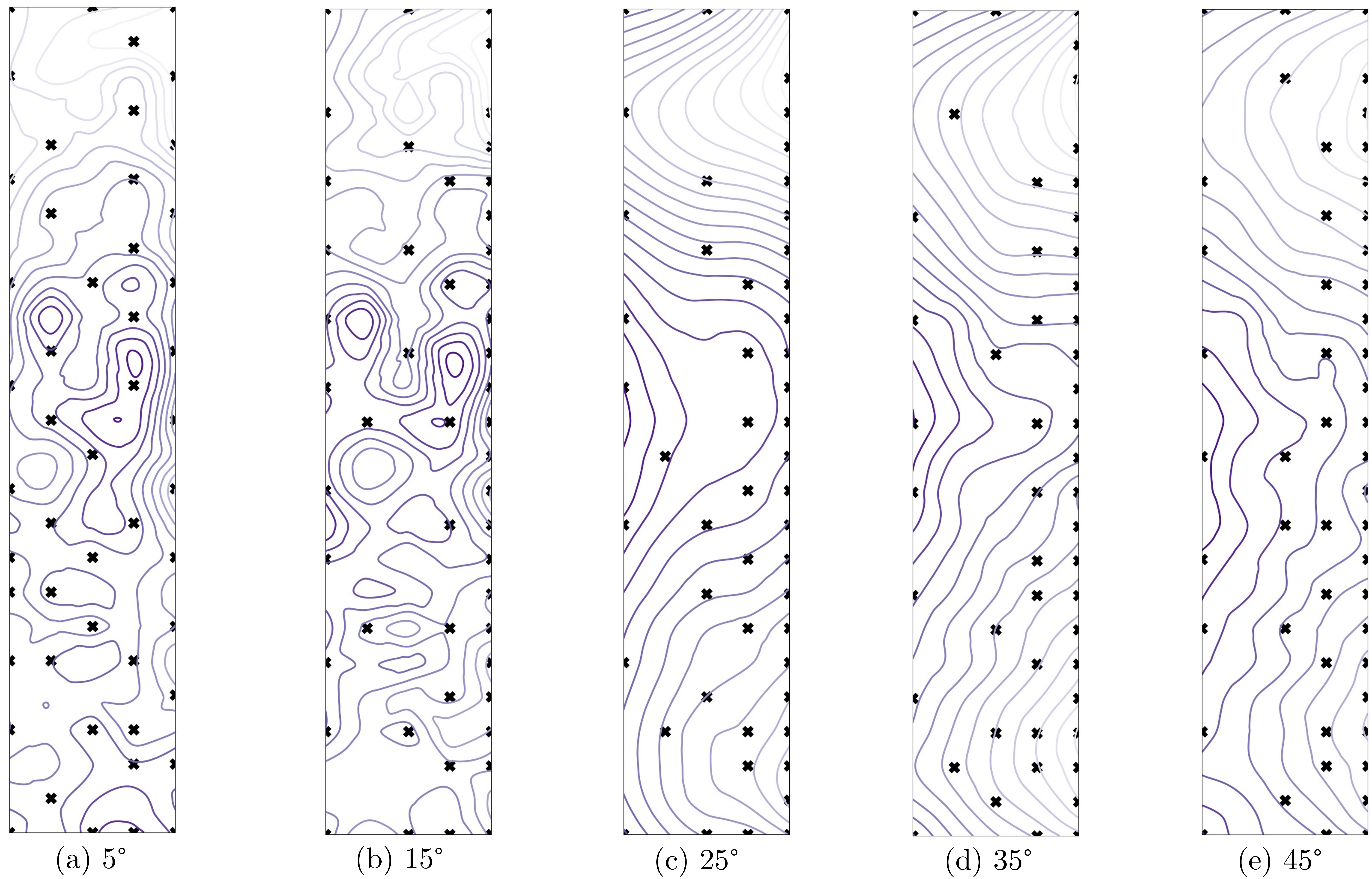}
    \caption{Optimal sensor locations: leeward scenario.}
    \label{fig:f14}
\end{figure}

\begin{figure}[H]
    \centering
    \includegraphics[width=0.9\textwidth]{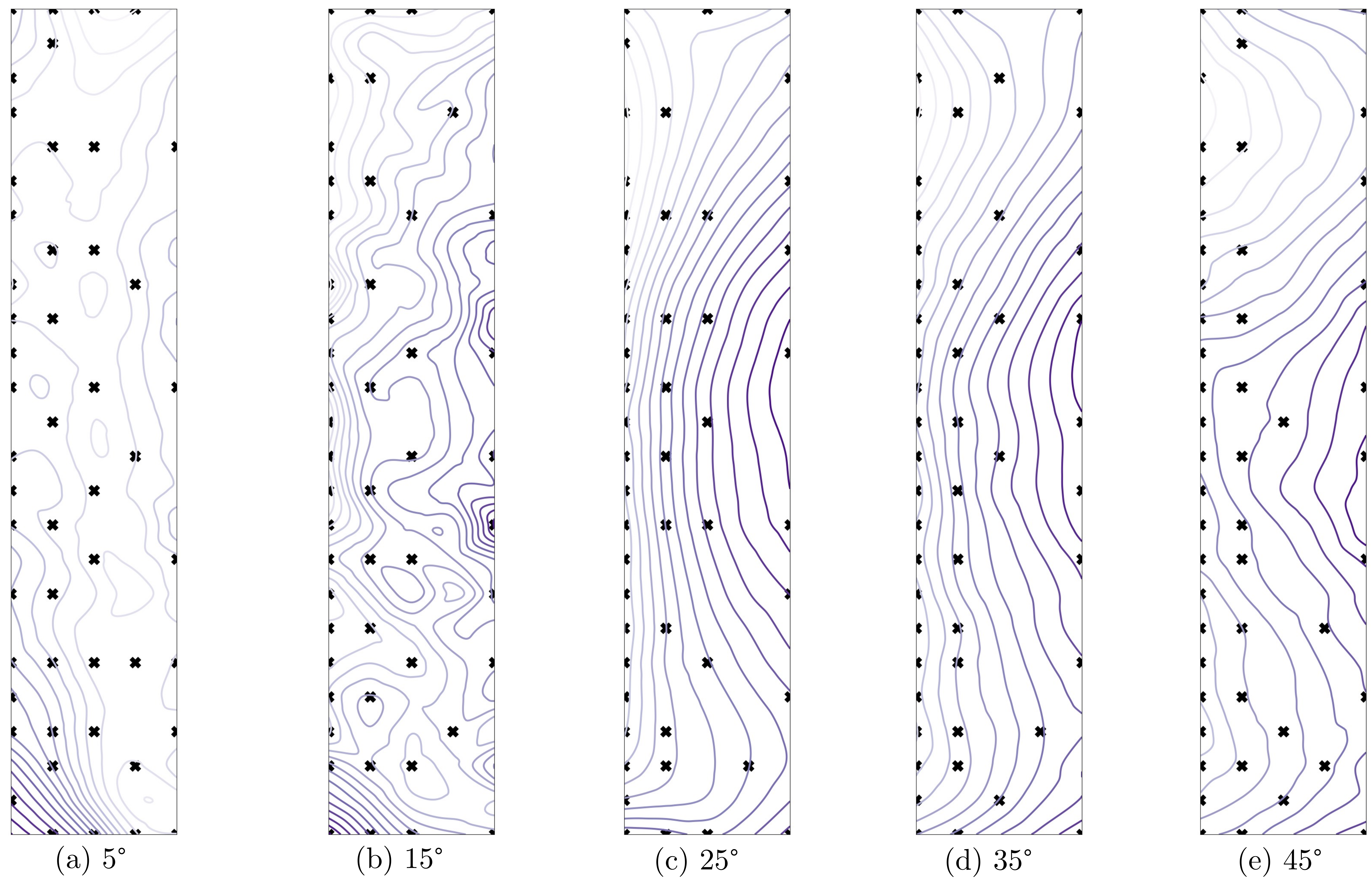}
    \caption{Optimal sensor locations: rightside scenario.}
    \label{fig:f15}
\end{figure}

\cref{fig:f12}-\cref{fig:f15} show optimal sensor placements for wind directions of $0^{\circ}$, $15^{\circ}$, $25^{\circ}$, $35^{\circ}$, and $45^{\circ}$. Clearly, contour lines on the windward surface's corner gradually become dense, indicating that wind pressures change dramatically, primarily due to fluid separation. As a result, many sensors are expected to be placed near corners and edges. Furthermore, in the $45^{\circ}$ case, contour lines gradually decrease from left to right. Because mean wind pressures gradually decrease, more sensors in the middle are required to provide a reliable reconstruction (See \cref{fig:f12} (e)). Meanwhile, the vortex appears when wind flows around the prism's side surface. The pressure pattern under a vortex generated by the windward corner and advected along the model's lateral face on the two side surfaces demonstrates the shift in the dense contour lines. Those dense contour lines indicate a strong fluid separation in the underlying area \cite{luo2021dynamic, carassale2012analysis}. It is worth noting that, despite the complex and dynamic nature of pressure field, the proposed sensor placement method is capable of capturing these intense regions of fluctuation and locating sensors to these positions.

\subsection{Computational effort}
\label{sec46}

\begin{figure}[H]
    \centering
    \includegraphics[width=0.95\textwidth]{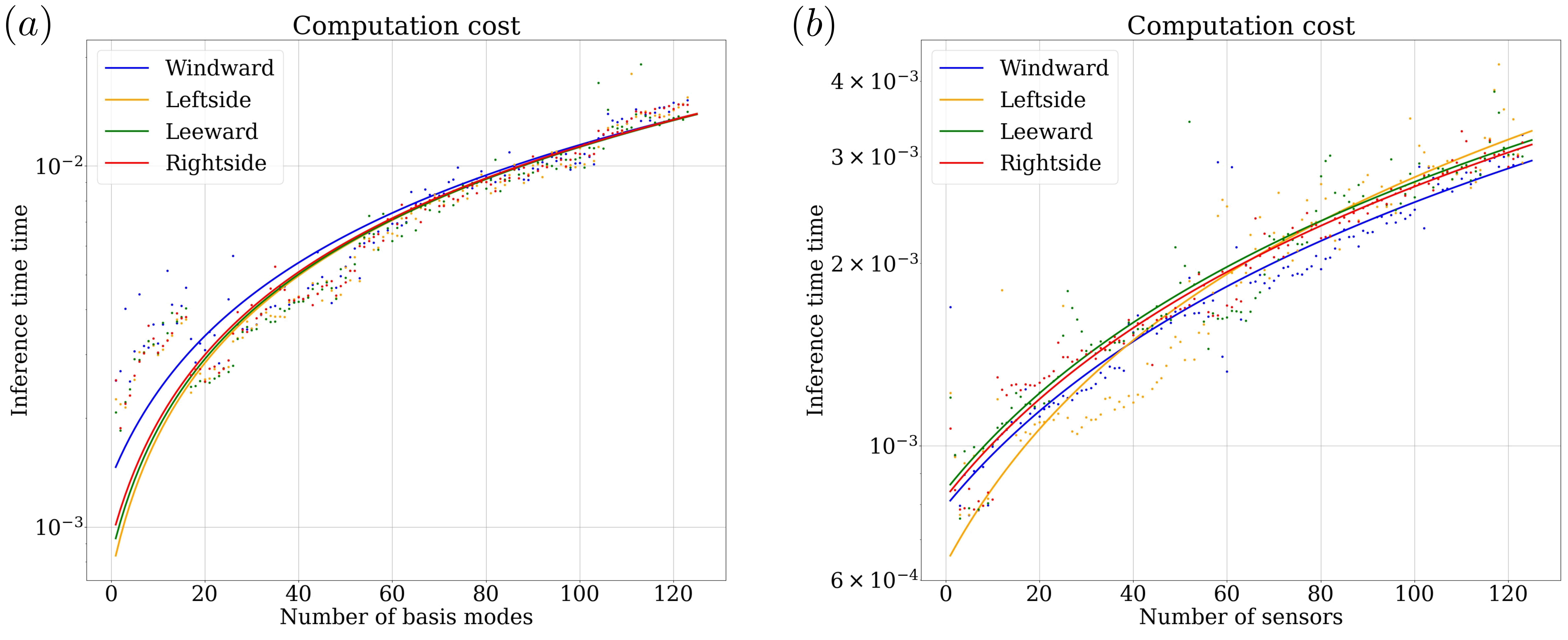}
    \caption{\textcolor{black}{Scalability analysis results: Each dot represents an experiment with a specific number of basis and sensors. The line plot is produced by fitting the wall clock computational times (dots) linearly using ordinary least squares optimization. Overall, the distribution of the various sets of dots agrees well with the fitted linear model. (a) fixed sensor number, varying basis number and (b) fixed basis number, varying sensor number}}
    \label{fig:f16}
\end{figure}

\textcolor{black}{In our experiments, pressure taps were evenly distributed on the surfaces of buildings, with 0.35 m row spacing and 5 columns on each face, resulting in 125 sensors for each building surface. A more dense set of pressure taps may be used in other wind tunnel experiments to measure both pressure dynamics and structural responses. We investigated the scalability of the proposed optimal sensor placement algorithm to determine its applicability to larger problems with orders of magnitude more pressure taps. We devised two sets of experiments in particular. First, we set the number of sensors to 125 and paired them with a range of basis numbers ranging from 1 to 124. Second, we set the number of basis to 124 and experimented with various sensor numbers. We used models trained with the aforementioned hyper-parameters to make inferences for the test data and recorded the computational cost in terms of wall-clock time.}
The scaling results for the various sensors/basis are shown in \cref{fig:f16}. 
\textcolor{black}{We fitted a linear model to these calculated computational costs to better present the scaling trend (See the line plots in the \cref{fig:f16}). This is accomplished through the use of ordinary least squares optimization.}
Results show that the model can achieve linear scaling on up to 125 sensors/basis. It should be noted that the SVD basis was used in these experiments. Even with full-rank sensors and basis, the total effort for training a model is less than a second, indicating that the proposed data-driven algorithm is both cost-effective and practical. For a generally larger problem such as three-dimensional turbulence, the identity basis may produce the lowest reconstruction error at a given number of sensors. This is because no empirical information is lost when constructing a low-rank approximation of the data. Still, by the same token, the identity basis can result in impractically long run times for a large data set \cite{manohar2018data}. In these cases, additional speedup can be obtained using a data-parallel implementation of the training algorithm, which distributes the batches across the cluster nodes and replicates the model \cite{yu2008mixed, wang2019visually}. 

\section{Concluding Remarks}
\label{sec5}
This study investigated data-driven sparse sensing approaches and adapted them to identify optimal sensor locations for reconstructing pressure field around a tall building. In order to determine the optimal sensor locations, the wind tunnel data was divided into three groups, i.e., training data, validation data, and testing data. The algorithm tailors a basis to the training data, which is then used in a QR decomposition to rank wind pressure taps by \textit{importance} based on reconstruction performance. This method allows for the removal and relocation of monitoring sensors with low information content as needed, as well as the provision of higher-ranking monitoring locations with higher-quality sensors or measuring at a higher frequency. This rank can also be used as a decision-making tool when looking for additional monitoring locations. A comparison of the reconstructed wind pressure data and the actual monitored wind pressure data reveals that the monitored wind pressure data can be satisfactorily reconstructed even with a limited number of sensors, indicating that the proposed method can serve as an efficient tool for reconstruction of random pressure fields around building. 

The proposed scheme is applicable to pressures in the vicinity of other geometries as well as low and mid-rise buildings. \textcolor{black}{Beyond that, we see the proposed algorithm as a reliable component in the design chain of the next generation of intelligent structures, where engineering elements such as sensing, actuating, and signal processing will be integrated into the overall system. Specifically, the rising scale and resolution of modern numerical simulations has resulted in a plethora of fluid-structure data. Although we may attain very high-level fidelity in the simulation, we are often limited in applications to a few noisy sensors. Therefore, we envision the use of large-scale CFD simulation data in conjunction with sensing techniques to develop a hybrid design and health monitoring framework for wind tunnel experiments. Utilizing deep learning, the optimal sensor location approach can be used to reconstruct the pressure field in real-time based on measurements from sparsely acquired data, offering an additional application.}

\section*{Acknowledgments}
This work is supported by the U.S. Department of Energy (DOE), Office of Science, Advanced Scientific Computing Research under Award Number DE-SC-0012704. \textcolor{black}{The authors also thank the anonymous reviewers for their comments and suggestions, which helped to improve the manuscript's quality and clarity.}

\bibliographystyle{elsarticle-num}
\bibliography{sample}

\begin{thebibliography}{10}
\expandafter\ifx\csname url\endcsname\relax
  \def\url#1{\texttt{#1}}\fi
\expandafter\ifx\csname urlprefix\endcsname\relax\def\urlprefix{URL }\fi
\expandafter\ifx\csname href\endcsname\relax
  \def\href#1#2{#2} \def\path#1{#1}\fi

\bibitem{kareem1992dynamic}
A.~Kareem, Dynamic response of high-rise buildings to stochastic wind loads,
  Journal of Wind Engineering and Industrial Aerodynamics 42~(1-3) (1992)
  1101--1112.

\bibitem{gu2004across}
M.~Gu, Y.~Quan, Across-wind loads of typical tall buildings, Journal of Wind
  Engineering and Industrial Aerodynamics 92~(13) (2004) 1147--1165.

\bibitem{kareem1985lateral}
A.~Kareem, Lateral-torsional motion of tall buildings to wind loads, Journal of
  Structural Engineering 111~(11) (1985) 2479--2496.

\bibitem{papadimitriou2004optimal}
C.~Papadimitriou, Optimal sensor placement methodology for parametric
  identification of structural systems, Journal of sound and vibration
  278~(4-5) (2004) 923--947.

\bibitem{meo2005optimal}
M.~Meo, G.~Zumpano, On the optimal sensor placement techniques for a bridge
  structure, Engineering structures 27~(10) (2005) 1488--1497.

\bibitem{jiang2007pseudospectra}
X.~Jiang, H.~Adeli, Pseudospectra, music, and dynamic wavelet neural network
  for damage detection of highrise buildings, International Journal for
  Numerical Methods in Engineering 71~(5) (2007) 606--629.

\bibitem{yi2011optimal}
T.-H. Yi, H.-N. Li, M.~Gu, Optimal sensor placement for structural health
  monitoring based on multiple optimization strategies, The Structural Design
  of Tall and Special Buildings 20~(7) (2011) 881--900.

\bibitem{tan2020computational}
Y.~Tan, L.~Zhang, Computational methodologies for optimal sensor placement in
  structural health monitoring: A review, Structural Health Monitoring 19~(4)
  (2020) 1287--1308.

\bibitem{yao1993sensor}
L.~Yao, W.~A. Sethares, D.~C. Kammer, Sensor placement for on-orbit modal
  identification via a genetic algorithm, AIAA journal 31~(10) (1993)
  1922--1928.

\bibitem{liu2008optimal}
W.~Liu, W.-c. Gao, Y.~Sun, M.-j. Xu, Optimal sensor placement for spatial
  lattice structure based on genetic algorithms, Journal of Sound and Vibration
  317~(1-2) (2008) 175--189.

\bibitem{yi2015optimal}
T.-H. Yi, H.-N. Li, G.~Song, X.-D. Zhang, Optimal sensor placement for health
  monitoring of high-rise structure using adaptive monkey algorithm, Structural
  Control and Health Monitoring 22~(4) (2015) 667--681.

\bibitem{manohar2018data}
K.~Manohar, B.~W. Brunton, J.~N. Kutz, S.~L. Brunton, Data-driven sparse sensor
  placement for reconstruction: Demonstrating the benefits of exploiting known
  patterns, IEEE Control Systems Magazine 38~(3) (2018) 63--86.

\bibitem{erichson2020shallow}
N.~B. Erichson, L.~Mathelin, Z.~Yao, S.~L. Brunton, M.~W. Mahoney, J.~N. Kutz,
  Shallow neural networks for fluid flow reconstruction with limited sensors,
  Proceedings of the Royal Society A 476~(2238) (2020) 20200097.

\bibitem{yang2021adaptive}
C.~Yang, An adaptive sensor placement algorithm for structural health
  monitoring based on multi-objective iterative optimization using weight
  factor updating, Mechanical Systems and Signal Processing 151 (2021) 107363.

\bibitem{brunton2022data}
S.~L. Brunton, J.~N. Kutz, Data-driven science and engineering: Machine
  learning, dynamical systems, and control, Cambridge University Press, 2022.

\bibitem{candes2006stable}
E.~J. Candes, J.~K. Romberg, T.~Tao, Stable signal recovery from incomplete and
  inaccurate measurements, Communications on Pure and Applied Mathematics: A
  Journal Issued by the Courant Institute of Mathematical Sciences 59~(8)
  (2006) 1207--1223.

\bibitem{donoho2006compressed}
D.~L. Donoho, Compressed sensing, IEEE Transactions on information theory
  52~(4) (2006) 1289--1306.

\bibitem{candes2008introduction}
E.~J. Cand{\`e}s, M.~B. Wakin, An introduction to compressive sampling, IEEE
  signal processing magazine 25~(2) (2008) 21--30.

\bibitem{candes2006robust}
E.~J. Cand{\`e}s, J.~Romberg, T.~Tao, Robust uncertainty principles: Exact
  signal reconstruction from highly incomplete frequency information, IEEE
  Transactions on information theory 52~(2) (2006) 489--509.

\bibitem{gilbert2010sparse}
A.~Gilbert, P.~Indyk, Sparse recovery using sparse matrices, Proceedings of the
  IEEE 98~(6) (2010) 937--947.

\bibitem{li2006very}
P.~Li, T.~J. Hastie, K.~W. Church, Very sparse random projections, in:
  Proceedings of the 12th ACM SIGKDD international conference on Knowledge
  discovery and data mining, 2006, pp. 287--296.

\bibitem{dasgupta2013experiments}
S.~Dasgupta, Experiments with random projection, arXiv preprint arXiv:1301.3849
  (2013).

\bibitem{fowler2009compressive}
J.~E. Fowler, Compressive-projection principal component analysis, IEEE
  transactions on image processing 18~(10) (2009) 2230--2242.

\bibitem{halko2011finding}
N.~Halko, P.-G. Martinsson, J.~A. Tropp, Finding structure with randomness:
  Probabilistic algorithms for constructing approximate matrix decompositions,
  SIAM review 53~(2) (2011) 217--288.

\bibitem{higham2000qr}
N.~J. Higham, Qr factorization with complete pivoting and accurate computation
  of the svd, Linear Algebra and its Applications 309~(1-3) (2000) 153--174.

\bibitem{TPU}
Tpu aerodynamic database: Wind pressure database based on wind tunnel
  experiment for high-rise building
  (http://wind.arch.t-kougei.ac.jp/system/eng/contents/code/tpu).

\bibitem{tamura1997proper}
Y.~Tamura, H.~Ueda, H.~Kikuchi, K.~Hibi, S.~Suganuma, B.~Bienkiewicz, Proper
  orthogonal decomposition study of approach wind-building pressure
  correlation, Journal of wind engineering and industrial aerodynamics 72
  (1997) 421--431.

\bibitem{zhao2017effects}
D.-X. Zhao, B.-J. He, Effects of architectural shapes on surface wind pressure
  distribution: case studies of oval-shaped tall buildings, Journal of Building
  Engineering 12 (2017) 219--228.

\bibitem{kareem1984pressure}
A.~Kareem, J.~Cermak, Pressure fluctuations on a square building model in
  boundary-layer flows, Journal of Wind Engineering and Industrial Aerodynamics
  16~(1) (1984) 17--41.

\bibitem{kim2021pressure}
B.~Kim, N.~Yuvaraj, K.~T. Tse, D.-E. Lee, G.~Hu, Pressure pattern recognition
  in buildings using an unsupervised machine-learning algorithm, Journal of
  Wind Engineering and Industrial Aerodynamics 214 (2021) 104629.

\bibitem{luo2021dynamic}
X.~Luo, A.~Kareem, Dynamic mode decomposition of random pressure fields over
  bluff bodies, Journal of Engineering Mechanics 147~(4) (2021) 04021007.

\bibitem{zhou2021higher}
L.~Zhou, K.~T. Tse, G.~Hu, Y.~Li, Higher order dynamic mode decomposition of
  wind pressures on square buildings, Journal of Wind Engineering and
  Industrial Aerodynamics 211 (2021) 104545.

\bibitem{taira2017modal}
K.~Taira, S.~L. Brunton, S.~T. Dawson, C.~W. Rowley, T.~Colonius, B.~J. McKeon,
  O.~T. Schmidt, S.~Gordeyev, V.~Theofilis, L.~S. Ukeiley, Modal analysis of
  fluid flows: An overview, Aiaa Journal 55~(12) (2017) 4013--4041.

\bibitem{meng2018sensitivity}
F.-Q. Meng, B.-J. He, J.~Zhu, D.-X. Zhao, A.~Darko, Z.-Q. Zhao, Sensitivity
  analysis of wind pressure coefficients on caarc standard tall buildings in
  cfd simulations, Journal of Building Engineering 16 (2018) 146--158.

\bibitem{carassale2012analysis}
L.~Carassale, Analysis of aerodynamic pressure measurements by dynamic coherent
  structures, Probabilistic engineering mechanics 28 (2012) 66--74.

\bibitem{yu2008mixed}
Z.~Yu, S.~Hoyos, B.~M. Sadler, Mixed-signal parallel compressed sensing and
  reception for cognitive radio, in: 2008 IEEE International Conference on
  Acoustics, Speech and Signal Processing, IEEE, 2008, pp. 3861--3864.

\bibitem{wang2019visually}
H.~Wang, D.~Xiao, M.~Li, Y.~Xiang, X.~Li, A visually secure image encryption
  scheme based on parallel compressive sensing, Signal Processing 155 (2019)
  218--232.

\end{thebibliography}

\end{document}